\documentclass[aps,prb,singlecolumn,eqsecnum,a4paper]{revtex4}
\usepackage{epsfig}
\usepackage{amssymb}
\usepackage{amsmath}
\usepackage{bm}
\usepackage{color}

\newcommand{\be}{\begin{equation}}
\newcommand{\ee}{\end{equation}}
\newcommand{\bea}{\begin{eqnarray}}
\newcommand{\eea}{\end{eqnarray}}
\newcommand{\up}{\uparrow}
\newcommand{\down}{\downarrow}
\def\LL{{\mathcal{L}}}
\def\nn{\nonumber\\}
\def\fr#1{(\ref{#1})}
\def\sgn{{\rm sgn}}

\def\a{\alpha}

\begin{document}
\title{Entanglement Entropies of the quarter filled Hubbard model} 
\author{Pasquale Calabrese}
\affiliation{Dipartimento di Fisica dell'Universit\`a di Pisa and INFN, 56127 Pisa, Italy}
\author{Fabian H.L. Essler}
\affiliation{The Rudolf Peierls Centre for Theoretical Physics, Oxford
University, Oxford OX1 3NP, UK}
\author{Andreas M. L\"auchli}
\affiliation{Institut f\"ur Theoretische Physik, Universit\"at Innsbruck, A-6020 Innsbruck, Austria}
\begin{abstract}
We study  R\'enyi and von Neumann entanglement entropies in the ground
state of the one dimensional quarter-filled Hubbard model with
periodic boundary conditions. We show that  they exhibit an unexpected
dependence on system size: for $L=4\ {\rm mod\ 8}$ the results are in  
agreement with expectations based on conformal field theory, while for
$L=0\ {\rm mod\ 8}$ additional contributions arise. We show that these
can be understood in terms of a ``shell-filling'' effect, and we develop a
conformal field theory approach to calculate the additional
contributions to the entropies. These analytic results are found to be
in excellent agreement with density matrix renormalisation group computations
for weak Hubbard interactions. We argue that for larger interactions
the presence of a marginal irrelevant operator in the spin sector
strongly affects the entropies at the finite sizes accessible numerically,
and we present an effective way to take them into account.  
\end{abstract}
\maketitle

\section{Introduction}

The Hubbard model is a central paradigm of strongly correlated
electron systems. Its 1D version has attracted much attention for
decades, because it is exactly solvable and exhibits a Mott metal to
insulator transition \cite{book}. The Hamiltonian for periodic
boundary conditions is
\be
H_{\rm Hubb}=-t\sum_{j=1}^L\sum_{\sigma=\uparrow,\downarrow}(c^\dagger_{j,\sigma}c_{j+1,\sigma}+ 
c^\dagger_{j+1,\sigma}c_{j,\sigma})
+U\sum_j n_{j,\uparrow}\ n_{j,\downarrow},
\label{HHubb}
\ee
where $c^\dagger_{j,\sigma}$ are fermionic spin-$\frac{1}{2}$ 
creation operators at site $j$ with spin $\sigma=\up,\down$,
$n_{j,\sigma}=c^\dagger_{j,\sigma}c_{j,\sigma}$, and we only consider
repulsive interactions $U\geq 0$. 
It is known from the exact solution that the ground state of \fr{HHubb} below half
filling (less than one fermion per site) is metallic and the low
energy physics of the model is described by a spin and charge
separated Luttinger liquid \cite{book}, equivalent to the semi-direct
product of two conformal field theories each with central charge $c=1$
\cite{FSspectrum} (so that the total central charge is $c=2$).
This correspondence has proved extremely useful in
characterizing the physical properties of the Hubbard model at low
energies. Over the last decade or so entanglement entropies (EEs) have
developed into a powerful tool for analysing many-body quantum
systems, in particular in relation to quantum criticality and
topological order \cite{rev}. In spite of this, a detailed analysis of the
EEs of the 1D Hubbard model beyond establishing the
leading behaviour\cite{CFT2} has been missing. 

In one dimensional systems described by conformal field theories (CFT)
in appropriate scaling limits very general predictions for the ground
state entanglement are known \cite{CFT0,CFT}. In order to be specific,
let us consider the ground state $|{\rm GS}\rangle$ of a finite,
periodic 1D system of length $L$ and partition the latter into a
finite block $A$ of length $\ell$ and its complement $\bar{A}$. The
density matrix of the entire system is then $\rho=|{\rm
  GS}\rangle\langle{\rm GS}|$, and we will denote the reduced density
matrix of block $A$ by $\rho_A\equiv {\rm Tr}_{\bar A} (\rho) $.  
Widely used measures of entanglement are the R\'enyi entropies 
\be
S_n = \frac{1}{1-n}\ln[{\rm Tr}{\rho_A^n}]\ .
\ee
They encode the full information on the spectrum of $\rho_A$
\cite{cl-08}, and in the limit $n\to 1$ reduce to the more widely
used von Neumann entropy  
\be 
S_1=-{\rm Tr}{\rho_A\ln\rho_A}.
\ee 
CFT predicts that the ground state EEs $S_n$ are given by
\be
S_{n} =\frac{c}{6}\left(1+\frac{1}{n}\right) \ln \Big(\frac{L}\pi
\sin\frac{\pi \ell}L\Big) +c_n+o\big(1\big)\,, 
\label{criticalent}
\ee
where $c$ is the central charge, $c_n$ are non-universal additive
constants, and $o(1)$ denotes terms that vanish for $\ell\to\infty$.  
The result \fr{criticalent} is valid beyond the scaling limit, i.e. applies
to lattice models underlying the CFT under consideration, as long as
subsystem size $\ell$ (and also $L-\ell$) is large compared to the
lattice spacing. The validity of \fr{criticalent} has been confirmed
for a large number of quantum spin-chains and models of interacting
electrons, see \cite{rev} for recent reviews. We note in passing that
the knowledge of the EEs has led to a deeper
understanding of numerical algorithms based on matrix product states
\cite{mps} and has aided the development of novel computational
methods \cite{cirac}.  

\par
In a recent short communication \cite{elc-13}, we have shown that the
ground state EEs for the Hubbard model do not
always follow (\ref{criticalent}). Ref. \cite{elc-13} focussed
on the particular case of a quarter-filled band, i.e. one electron per two sites
$N_\uparrow=N_\downarrow={L}/{4}$, although our findings generalize to
other fillings and, in fact, to other models (indeed the same effect was probably  
present also in Ref. \cite{bt-14}).

\begin{figure}[ht]
\begin{center}
\includegraphics[width=0.6\linewidth]{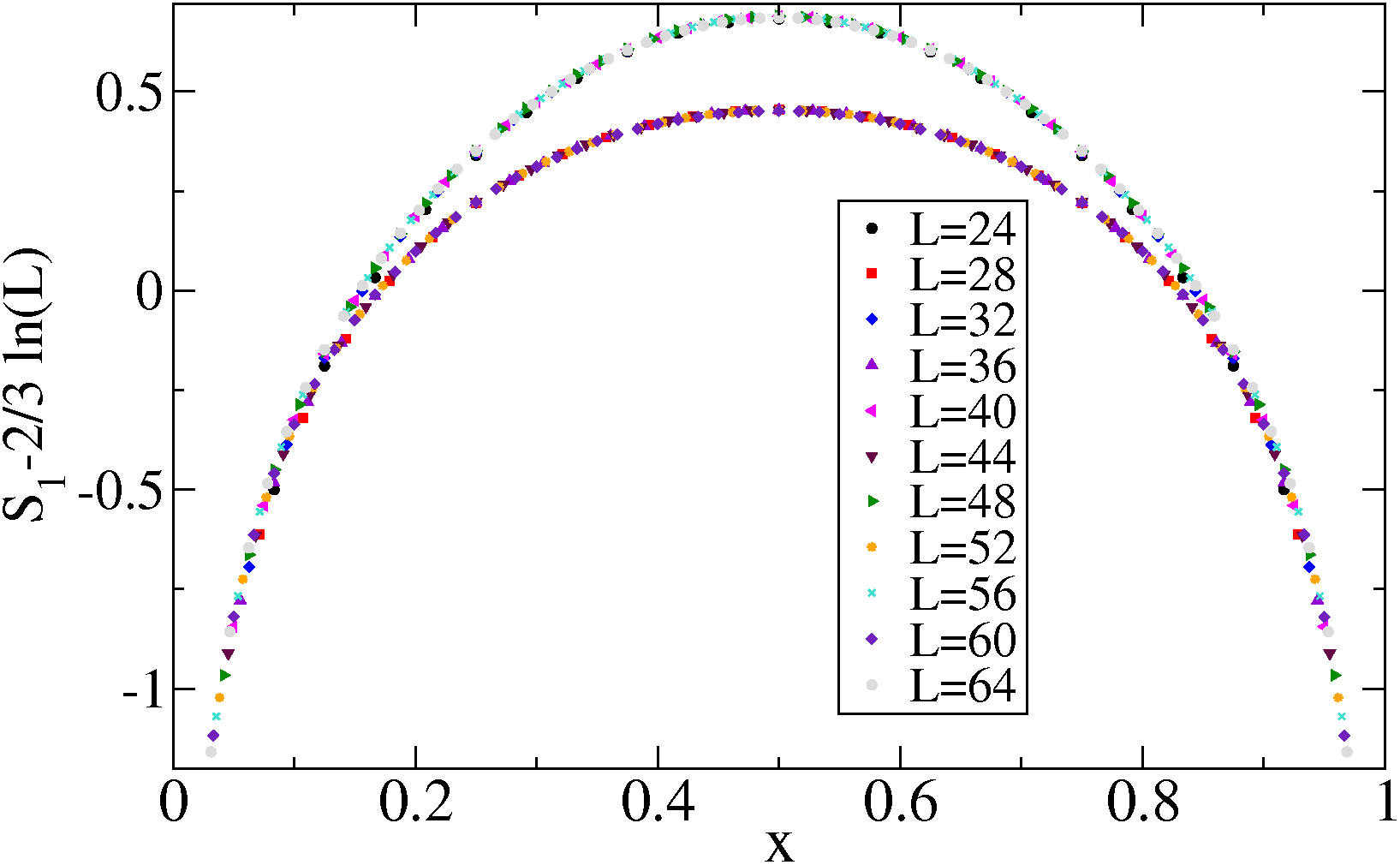}
\end{center}
\caption{DMRG data for the bipartite entanglement entropy in  in the
ground state of the Hubbard chain for $U=t$. We report $S_1-2/3 \ln
L$ as a function of the subsystem size $x=\ell/L$  for
$L=24,28,32,36,40,44,48,52,56,60,64$. The lower and upper branches
corresponds to lattice lengths $L=4\ {\rm mod}\ 8$ and $L=0\ {\rm
  mod}\ 8$ respectively.} 
\label{fig:vN1s}
\end{figure}

The main result of Ref. \cite{elc-13} is summarized in
Fig. \ref{fig:vN1s}, where we plot the subtracted entanglement
entropies $S_1-2/3 \ln L$ for a quarter-filled Hubbard model at $U=t$
for a number of different lattice lengths $L$. Interestingly, both the
$L=4\ {\rm  mod}\ 8$ and the $L=0\ {\rm   mod}\ 8$ data exhibit
scaling collapse, \emph{but to different functions}. 
The entropy for lattice lengths $L=4\ {\rm mod}\ 8$ is well-described
by the CFT result \fr{criticalent} with $n=1$, while for $L=0\ {\rm
  mod}\ 8$ there is an additional positive
contribution. Interestingly, the latter can also be obtained by means
of CFT \cite{elc-13}.  
The physical origin of this unusual behaviour can be traced back to a
shell filling effect. 

In this manuscript we continue our study of entanglement entropies
in the 1D Hubbard model and our goal is twofold. On the one hand, we
report details of the explanation and derivation of the shell filling
effect in Fig. \ref{fig:vN1s}, which were not given previously in
Ref.~[\onlinecite{elc-13}]. On the other hand, we present for the
first time the results of our extensive numerical analysis of various 
entanglement entropies and Hubbard couplings. The manuscript is
organized as follows. In Sec. \ref{tbm} we explain the shell filling
effect for the non-interacting case $U=0$ and develop a CFT
description of the low-energy degrees of freedom for periodic boundary
conditions and different system sizes.
In Sec. \ref{sec:hm} we review elements of the exact solution of the
Hubbard model, and in Sec. \ref{sec:LL} we relate them to Luttinger
liquid theory. In Sec. \ref{sec:CFT} we report the CFT derivation of
the entanglement entropies explaining, among the other things,   
the results in Fig. \ref{fig:vN1s}. Our analytic predictions are
compared to numerical results in Sec. \ref{sec:num}, and deviations
are carefully analyzed. Finally in Sec. \ref{concl} we draw our
conclusions and discuss some open issues.

\section{Tight-Binding Model}
\label{tbm}
In order to understand the shell-filling effect in the ground state of
the Hubbard Hamiltonian it is instructive to first consider the non
interacting limit ($U=0$), which is the 1D tight-binding model for
spin-1/2 fermions 
\be
H_0=-t\sum_{j=1}^L\sum_{\sigma=\uparrow,\downarrow}(c^\dagger_{j,\sigma}c_{j+1,\sigma} c^\dagger_{j+1,\sigma}c_{j,\sigma}).
\ee
$H_0$ is diagonalized by going to momentum space
\be
c^\dagger_\sigma(p)=\frac{1}{\sqrt{L}}\sum_{j=1}^L e^{-ipj}c^\dagger_{j,\sigma}.
\ee
Imposing periodic boundary conditions leads to the quantization of the
momenta
\be
p_m=\frac{2\pi m}{L} ,\qquad -\frac{L}{2}\leq m<\frac{L}{2}.
\ee
We are interested a quarter filled band and zero magnetization, i.e.
\be
N_\uparrow=N_\downarrow={L}/{4}\ .
\ee
\subsection{Ground State}
It is now straightforward to see that the precise structure of the
ground state depends sensitively on whether $N_\uparrow=N_\downarrow$
are even or odd. For later convenience we introduce the notation
\be 
|N_\uparrow,N_\downarrow;L\rangle_{\rm GS}
\ee
for the ground state of the Hamiltonian in the sector with $N_\sigma$ particles
with spin $\sigma$ for a chain of length $L$.
\begin{enumerate}
\item{} Let us first consider $N_\sigma$ to be odd, i.e. $N_\uparrow=N_\downarrow=2n+1$.
In this case the quarter-filled ground state is unique and given by the
{\it symmetric} Fermi sea     
\be
|2n+1,2n+1;8n+4\rangle_{\rm GS}= \prod_{m=-n}^n
c^\dagger_\uparrow(p_m)c^\dagger_\downarrow(p_m)|0\rangle\ ,
\label{gsbad}
\ee
where  $|0\rangle$ is the fermionic vacuum state.

\item{} When $N_\sigma=L/4=2n$ is even, the Fermi sea for a given spin
species is necessarily \emph{asymmetric}. As a consequence the ground
state is not unique. The simplest way to construct the degenerate
ground states is to start from the unique ground state of the
Hamiltonian with $L=8n$, but with one particle less per species,
i.e. $N_\sigma=L/4-1$. As $N_\sigma=L/4-1$ is now odd, the ground
state is the symmetric Fermi sea
\be
|2n-1,2n-1;8n\rangle_{\rm GS}=
\prod_{m=-L/8+1}^{L/8-1}c^\dagger_\uparrow(p_m)c^\dagger_\downarrow(p_m) |0\rangle.
\label{GSm1}
\ee
The corresponding Fermi momentum is given by
\be
k_F=\frac{\pi}{4}=p_{{L}/{8}},
\ee
so that in the grand canonical ensemble the single-particle energy is
\be
\epsilon_\sigma(k)=-2t\left[\cos(k)-\cos(k_F)\right].
\ee
As $\epsilon_\sigma(\pm k_F)=0$ there are four zero modes, which give
rise to a total of \emph{sixteen} degenerate ground states. Four of
these occur at exactly quarter filling $N_\uparrow=N_\downarrow={L}/4=2n$
\bea
\label{4states}
|2n,2n;8n\rangle_{\rm GS,1}&=&
c^\dagger_\uparrow(k_F)\ c^\dagger_\downarrow(-k_F)
\left|2n-1,2n-1;8n\right\rangle_{\rm GS}\ ,\nn
|2n,2n;8n\rangle_{\rm GS,2}&=&
c^\dagger_\downarrow(k_F)\ c^\dagger_\uparrow(-k_F)
\left|2n-1,2n-1;8n\right\rangle_{\rm GS},\nn
|2n,2n;8n\rangle_{\rm GS,3}&=&
c^\dagger_\uparrow(k_F)\ c^\dagger_\downarrow(k_F)
\left|2n-1,2n-1;8n\right\rangle_{\rm GS}\ ,\nn
|2n,2n;8n\rangle_{\rm GS,4}&=&
c^\dagger_\downarrow(-k_F)\ c^\dagger_\uparrow(-k_F)
\left|2n-1,2n-1;8n\right\rangle_{\rm GS}.
\eea

States $|2n,2n;8n\rangle_{\rm GS,3}$ and $|2n,2n;8n\rangle_{\rm GS,4}$
have momenta $\pm 2k_F$ respectively, while $|2n,2n;8n\rangle_{\rm
  GS,1}$ and $|2n,2n;8n\rangle_{\rm GS,2}$ have momentum zero. 
The Hubbard interaction splits these degeneracies and selects a unique
ground state in the sector with $N_\uparrow=N_\downarrow=2n$. In the limit
$U\to 0$, the Hubbard model ground state reduces to
\be
|+ \rangle=\frac{1}{\sqrt{2}}\Big[
|2n,2n;8n\rangle_{\rm GS,1}+|2n,2n;8n\rangle_{\rm GS,2}\Big]\ .
\label{desired}
\ee
\end{enumerate}
\subsection{Finite-size spectrum}
In order to make contact with CFT, it is useful to have in hand
expressions for the finite-size spectra of low-lying excited
states. This is straightforward since we are dealing with a free
fermionic theory.

\subsubsection{$L=8n+4$}

After some calculations we find
\bea
E(\Delta N_\sigma,D_\sigma,M_\sigma^{\pm})=Le_0-\frac{\pi
  v}{3L}+\frac{2\pi v}{L}\sum_{\sigma}
\left[\left(\frac{\Delta N_\sigma}{2}\right)^2+
\big(D_\sigma\big)^2\right]+\frac{2\pi v}{L}\sum_{\sigma}\sum_{n>0}
n\left[M_{n,\sigma}^++M_{n,\sigma}^-\right],
\label{energy}
\eea
where $v=\epsilon'(k_F)$, $e_0$ is the ground state energy density in
the thermodynamic limit, $M_{n,\sigma}^\pm$ are integers and
\be
\Delta N_\sigma=N_\sigma-\frac{L}{4}\ ,\qquad
D_\sigma=
\bigg\{\begin{array}{l l}
{\rm integer} &\text{if}\ \Delta N_\sigma\ \text{is even}\\
{\rm half-odd\ integer} &\text{if}\ \Delta N_\sigma\ \text{is odd}
\end{array}
\quad\sigma=\uparrow,\downarrow .
\ee
This means that the unique ground state is obtained by taking
$N_\sigma=L/4$ and $D_\sigma=0$, which gives the standard ``conformal''
result for the ground state energy
\be
E_{\rm GS}=Le_0-\frac{\pi v}{3L}+o(L^{-1}).
\ee
The momentum of these states is
\be
P(\Delta N_\sigma,D_\sigma,M_\sigma^{\pm})=
\sum_\sigma\frac{2\pi}{L}\left[\Delta N_\sigma D_\sigma
+\sum_{n>0}n\left(M^+_{n,\sigma}-M^-_{n,\sigma}\right)
\right]+\sum_\sigma
2k_FD_\sigma.
\label{mtm}
\ee
The spectra \fr{energy} and \fr{mtm} are of the same form as for a
compactified free boson.

\subsubsection{$L=8n$}

In this case we obtain the same expressions \fr{energy} and \fr{mtm}
for energy and momentum of low-lying excitations, but now the
``quantum numbers'' $D_\sigma$ take the values
\be
\Delta N_\sigma=N_\sigma-\frac{L}{4}\ ,\qquad
D_\sigma=
\bigg\{\begin{array}{l l}
{\rm integer} &\text{if}\ \Delta N_\sigma\ \text{is odd}\\
{\rm half-odd\ integer} &\text{if}\ \Delta N_\sigma\ \text{is even}
\end{array}
\quad\sigma=\uparrow,\downarrow .
\label{D8n}
\ee
As discussed before, the ground state is sixteenfold degenerate in
this case
\bea
\text{4 states:  } \Delta N_\sigma&=&0\ ,\quad D_\sigma=\pm\frac{1}{2}\ ,\nn
\text{8 states:  } \Delta N_\uparrow&=&\pm 1\ ,\Delta N_\downarrow=0\ ,\quad 
D_\uparrow=0\ ,D_\downarrow=\pm\frac{1}{2}\ , \text{and }
\uparrow\leftrightarrow\downarrow\ , \nn
\text{4 states:  } \Delta N_\sigma&=&\pm 1\ ,\quad D_\sigma=0.
\eea

\subsection{Correlation Functions of Local Operators for $L=8n$}
In the case $L=8n+4$ multi-point correlation functions of fermion
operators can be easily calculated using Wick's theorem. The 2-point
function is
\be
{}_{\rm GS}\langle 2n+1,2n+1;8n+4|
c^\dagger_{j,\sigma}c_{l,\tau}|2n+1,2n+1;8n+4\rangle_{\rm GS}
=\delta_{\sigma,\tau}\frac{\sin\big(k_F(j-l)\big)}
{L\sin\big(\frac{\pi}{L}(j-l)\big)}, 
\ee
where $k_F=\pi/4$. The case $L=8n$ is more complicated. Our aim is to
determine correlation functions in the state \fr{desired}. To that end
it is useful to first consider correlators on a chain of length $L=8n$
with two particles below quarter filling. In this case the
symmetrically filled Fermi sea \fr{GSm1} is the unique ground-state,
and we have
\be
{}_{\rm GS}\langle 2n-1,2n-1;8n|
c^\dagger_{j,\sigma}c_{l,\tau}|2n-1,2n-1;8n\rangle_{\rm GS}
=\delta_{\sigma,\tau}\frac{\sin\big(\tilde{k}_F(j-l)\big)}
{L\sin\big(\frac{\pi}{L}(j-l)\big)}, 
\ee
where
\be
\tilde{k}_F=\frac{\pi}{4}-\frac{\pi}{L}.
\label{ktilde}
\ee
As we are dealing with symmetrically filled Fermi seas, the result is
the same as for ground state correlators on lattices of length $L=8n+4$ with the replacement
\be
\tilde{k}_F\longrightarrow k_F=\frac{\pi}{4}.
\ee
Correlators with respect to the state $|+\rangle$ \fr{desired} can now
be worked out by first using the representation \fr{4states}, and then
applying Wick's theorem (in the state $|2n-1,2n-1;8n\rangle_{\rm
  GS}$). Let us denote the Green's function in the state $|+\rangle$
on a lattice with $L=8n$ sites by
\be
D_{\sigma,\tau}(j-l,L,\tilde{k}_F)=
\langle + |c^\dagger_{j,\sigma}c_{l,\tau}|+\rangle\ .
\ee
Let us now consider the Green's function in the following
excited state for a lattice of $L=8n+4$ sites
\be
|{\rm EXC}\rangle=\frac{1}{\sqrt{2}}\left[
c^\dagger_\up(k)c^\dagger_\down(-k)+
c^\dagger_\down(k)c^\dagger_\up(-k)
\right]\left|2n+1,2n+1;8n+4\right\rangle_{\rm GS},
\label{exc1}
\ee
where $k$ corresponds to the lower energy empty mode, i.e.
\be
k=k_F+\frac{\pi}{L}\ .
\ee
Clearly we have
\be
\langle {\rm EXC} |c^\dagger_{j,\sigma}c_{l,\tau}| {\rm EXC} \rangle= D_{\sigma,\tau}(j-l,L,{k}_F).
\ee
\emph{This means that rather than working with the ground-state for $L=8n$ we can work with
a particular excited state for $L=8n+4$ and in the end of our calculations replace $\tilde{k}_F$ by $k_F$}. 
As we are going to carry out a field theory calculation in
which $k_F$ is an external parameter this will be straightforward to do.
\subsection{Field Theory Description for the noninteracting model}
Let us now think of the quarter-filled tight-binding model in the thermodynamic
limit, where we have approached the latter for values of $L$ and $N_\sigma$ such
that we are dealing with a unique ground state. A naive linearization
of the fermion dispersion around the Fermi momentum gives rise to a
field theory description in terms of massless Dirac fermions 
\be
c_{j,\sigma}\sim\sqrt{a_0}\left[e^{ik_Fx/a_0}R_\sigma(x)+e^{-ik_Fx/a_0}L_\sigma(x)\right],
\ee
where $k_F={\pi}/{4}$ and $a_0$ represents the lattice spacing.
Here $R_\sigma(x)$ describes the Fourier modes of $c_{j,\sigma}$ with
momenta close to $k_F$, while $L_\sigma(x)$ incorporates the modes in the
vicinity of $-k_F$. In terms of these fields, the Hamiltonian reads
\be
H=iv\sum_{\sigma=\uparrow,\downarrow}\int dx\left[L^\dagger_{\sigma}(x)\partial_x
  L_{\sigma}(x)-R_{\sigma}^\dagger(x)\partial_x R_{\sigma}(x) \right],
\ee
where $v=2ta_0\sin(k_F)$ is the Fermi velocity.
Periodic boundary conditions on the lattice fermions $c_{L+1}=c_1$
imply that for lattice lengths $L=8n+4$
\be
R_\sigma(x+La_0)=-R_\sigma(x)\ ,\quad
L_\sigma(x+La_0)=-L_\sigma(x)\ .
\label{bcFermi8n+4}
\ee
On the other hand, for $L=8n$ we have
\be
R_\sigma(x+La_0)=R_\sigma(x)\ ,\quad
L_\sigma(x+La_0)=L_\sigma(x)\ .
\label{bcFermi8n}
\ee
\subsubsection{Bosonization}
The Fermi fields can be bosonized\cite{affleck} using
\be
R_\sigma(x)\sim \frac{\eta_\sigma}{\sqrt{La_0}}\ :e^{i\sqrt{4\pi}\varphi_\sigma(x)}:\ , \qquad 
L_\sigma(x)\sim \frac{\eta_\sigma}{\sqrt{La_0}}\ :e^{-i\sqrt{4\pi}\bar{\varphi}_\sigma(x)}:\ ,
\label{bosonization}
\ee
where $\eta_\sigma$ are Klein factors. In terms of the bosonic fields
the low-energy effective Hamiltonian is 
\be
H=\sum_{\sigma=\uparrow,\downarrow}\frac{v}{2}\int dx \left[(\partial_x\Phi_\sigma)^2+
(\partial_x\Theta_\sigma)^2\right],
\label{freeboson}
\ee
where we have defined canonical Bose fields $\Phi_\sigma$ and their
dual fields $\Theta_\sigma$ by
\be
\Phi_\sigma=\varphi_\sigma+\bar{\varphi}_\sigma\ ,\qquad
\Theta_\sigma=\varphi_\sigma-\bar{\varphi}_\sigma\ .
\ee
Instead for working with ``up'' and ``down'' Bose fields, we can
alternatively work with spin and charge bosons 
\bea
\varphi_c&=&\frac{\varphi_\uparrow+\varphi_\downarrow}{\sqrt{2}}\ ,\quad
\bar{\varphi}_c=\frac{\bar{\varphi}_\uparrow+\bar{\varphi}_\downarrow}{\sqrt{2}}\ ,\qquad
\Phi_c=\varphi_c+\bar{\varphi}_c\ ,\quad
\Theta_c=\varphi_c-\bar{\varphi}_c\ ,
\nn
\varphi_s&=&\frac{\varphi_\uparrow-\varphi_\downarrow}{\sqrt{2}}\ ,\quad
\bar{\varphi}_s=\frac{\bar{\varphi}_\uparrow-\bar{\varphi}_\downarrow}{\sqrt{2}}\  ,\qquad
\Phi_s=\varphi_s+\bar{\varphi}_s\ ,\quad
\Theta_s=\varphi_s-\bar{\varphi}_s\ .
\eea
In terms of these the Hamiltonian reads
\be
H=\sum_{\alpha=c,s}\frac{v}{2}\int dx \left[(\partial_x\Phi_\alpha)^2+
(\partial_x\Theta_\alpha)^2\right].
\ee
The normalization of $H$ is such that, in the infinite
volume limit, the two-point functions of vertex operators are 
normalized as  
\be
\langle 0|
e^{i\beta\Phi_\alpha(x)}e^{-i\beta'\Phi_\alpha(0)}|0\rangle\sim
\delta_{\beta,\beta'}
\left[\frac{ia_0}{x}\right]^{\frac{\beta\beta'}{2\pi}}.
\label{norm1}
\ee

\subsubsection{Finite-Size Spectrum an Correlations for $L=8n+4$}
\label{Cor8n4}
In order to impose boundary conditions and work out the finite-size
energy spectrum, we recall the mode expansions of the chiral Bose fields
\bea
\varphi_\sigma(x,t)&=&\varphi_{\sigma,0}+\frac{x-vt}{\LL}Q_\sigma
+\sum_{n=1}^\infty\frac{1}{\sqrt{4\pi n}}\left[
e^{i\frac{2\pi n}{\LL}(x-vt)}a_{\sigma,R,n}+e^{-i\frac{2\pi
    n}{\LL}(x-vt)}a^\dagger_{\sigma,R,n}\right] ,\nn
\bar{\varphi}_\sigma(x,t)&=&\bar{\varphi}_{\sigma,0}+\frac{x+vt}{\LL}
\bar{Q}_\sigma+\sum_{n=1}^\infty\frac{1}{\sqrt{4\pi n}}\left[  
e^{-i\frac{2\pi n}{\LL}(x+vt)}a_{\sigma,L,n}+e^{i\frac{2\pi
    n}{\LL}(x+vt)}a^\dagger_{\sigma,L,n}\right] .
\eea
Here $\LL=La_0$ in the physical length of the system, while
$\varphi_{\sigma,0}$ and $Q_\sigma$ are zero momentum modes satisfying
canonical commutation relations
\be
[\varphi_{\sigma,0},Q_\sigma]=-\frac{i}{2}=
-[\bar{\varphi}_{\sigma,0},\bar{Q}_\sigma].
\ee
It is straightforward to check that the chiral Bose field fulfil the
following equal time commutation relations
\be
[\varphi_\sigma(x),\varphi_\sigma(y)]=
-[\bar{\varphi}_\sigma(x),\bar{\varphi}_\sigma(y)]=\frac{i}{4}\sgn(x-y)\ ,\qquad 
\lbrack\varphi_\sigma(x),\bar{\varphi}_\sigma(y)]=\frac{i}{4}\ .
\ee
We are now in a position to impose the boundary conditions
\fr{bcFermi8n+4}, which lead to the requirement that
$
e^{i\sqrt{4\pi}\varphi_\sigma(\LL)}=-e^{i\sqrt{4\pi}\varphi_\sigma(0)}$. Substituting
the mode expansions one obtains the following quantization conditions for the
zero modes
\be
e^{i\sqrt{4\pi}Q_\sigma}=1=e^{-i\sqrt{4\pi}\bar{Q}_\sigma}.
\ee
Hence the spectra of $Q_\sigma$ and $\bar{Q}_\sigma$ are
\be
q_{\sigma,m}=\sqrt{\pi}m\ ,\quad
\bar{q}_{\sigma,m'}=\sqrt{\pi}m'\ ,\quad m,m'\in\mathbb{Z}.
\label{quant1}
\ee
The corresponding boundary conditions on the Bose fields $\varphi_\sigma(x)$,
$\bar{\varphi}_\sigma(x)$ are
\bea
\varphi_\sigma(x+\LL)&=&\varphi_\sigma(x)+\sqrt{\pi}m_\sigma\ ,\nn
\bar{\varphi}_\sigma(x+\LL)&=&\bar{\varphi}_\sigma(x)+\sqrt{\pi}m'_\sigma\ .
\label{goodBC1}
\eea
Recalling that $\Phi_\sigma$ is a compactified field
\be
\Phi_\sigma(x)=\Phi_\sigma(x)+\sqrt{\pi}\ ,
\ee
we conclude that \fr{goodBC1} in fact correspond to periodic boundary
conditions on $\Phi_\sigma$. Substituting the mode expansions into the
expression for the Hamiltonian gives
\bea
H&=&\frac{v}{\LL}\left[\sum_\sigma Q_\sigma^2+\bar{Q}_\sigma^2+
\sum_\sigma\sum_{n>0}2\pi n\left(a^\dagger_{\sigma,R,n}a_{\sigma,R,n}
+a^\dagger_{\sigma,L,n}a_{\sigma,L,n}\right)\right].
\eea
This results in a finite-size spectrum of the form
\bea
E&=&\frac{2\pi v}{\LL}\left[\sum_\sigma \frac{\left(m_\sigma\right)^2
+\left(m'_\sigma\right)^2}{2}+
\sum_\sigma\sum_{n>0}n\left[M_{n,\sigma}^++M_{n,\sigma}^-\right]
\right].
\label{FSSboso}
\eea
Defining new quantum numbers
\be
\Delta N_\sigma=m_{\sigma}+m'_{\sigma}\ ,\quad
D_\sigma=\frac{m_{\sigma}-m'_{\sigma}}{2}\ ,
\label{ND}
\ee
we can rewrite the expression for the energy as
\bea
E&=&\frac{2\pi v}{\LL}\left[\sum_\sigma \left(\frac{\Delta N_\sigma}{2}\right)^2
+D_\sigma^2+
\sum_{n>0}n\left[M_{n,\sigma}^++M_{n,\sigma}^-\right]\right].
\label{E8n4}
\eea
As required this reproduces the lattice result \fr{energy}. Finally,
we wish to calculate the Green's function using the field theory
formalism. The ground state $|0\rangle$ is characterized by the zero
mode quantum numbers $m=m'=0$. Using the mode expansion,
a straightforward calculation gives
\be
\langle 0|:e^{-i\sqrt{4\pi}\varphi_\sigma(x)}:\ 
:e^{i\sqrt{4\pi}\varphi_\sigma(0)}:|0\rangle=\frac{i}{2\sin(\pi x/\LL)}.
\ee
Going back to the the bosonization identities \fr{bosonization}, we
conclude that
\bea
\langle R^\dagger_\sigma(x)R_\sigma(0)\rangle&=&\frac{i}{2\LL\sin(\pi
  x/\LL)}\ ,\nn
\langle L^\dagger_\sigma(x)L_\sigma(0)\rangle&=&-\frac{i}{2\LL\sin(\pi x/\LL)}.
\eea
The asymptotics of the lattice correlators follows then to be
\be
\langle c^\dagger_{j+\ell,\sigma}
c_{j,\sigma}\rangle\sim a_0\Big[e^{-ik_F\ell}
\langle R^\dagger_\sigma(\ell a_0)R_\sigma(0)\rangle +e^{ik_F\ell}
\langle L^\dagger_\sigma(\ell a_0)L_\sigma(0)\rangle\Big]
=\frac{\sin(k_F \ell)}{L\sin(\pi \ell/L)},
\ee
which, as required, agrees with the direct lattice calculation.
\subsubsection{Finite-Size Spectrum and Correlations for $L=8n$}
Following through the same steps leading to Eq. \fr{quant1}, we find
that for $L=8n$ the eigenvalues of the zero mode operators are now
given by
\be
q_{\sigma,m}=\sqrt{\pi}\big(m-\frac{1}{2}\big)\ ,\quad
\bar{q}_{\sigma,m}=\sqrt{\pi}\big(m-\frac{1}{2}\big),\quad
m,m'\in\mathbb{Z}.
\label{quant2}
\ee
These quantization conditions correspond to the following boundary
conditions on the Bose fields $\varphi_\sigma(x)$, $\bar{\varphi}_\sigma(x)$
\bea
\varphi_\sigma(x+\LL)&=&\varphi_\sigma(x)+\sqrt{\pi}\left[m_\sigma-\frac{1}{2}\right] ,\nn
\bar{\varphi}_\sigma(x+\LL)&=&\bar{\varphi}_\sigma(x)+\sqrt{\pi}\left[m'_\sigma-\frac{1}{2}
\right].
\label{twistedBCs}
\eea
The resulting energy spectrum is of the form
\bea
E&=&\frac{2\pi v}{L}\Big[\sum_\sigma \frac{\left(m_\sigma-\frac{1}{2}\right)^2}{2}
+\frac{\left(m'_\sigma-\frac{1}{2}\right)^2}{2}
+\sum_\sigma\sum_{n>0}n\left[M_{n,\sigma}^++M_{n,\sigma}^-\right]
\Big].
\eea
Using the definitions \fr{ND} we can bring this to the same form
\fr{E8n4} as for $L=8n+4$, but now with a difference in the allowed
values of $D_\sigma$. This again agrees with the lattice result \fr{D8n}.
Our particular states of interest $|2n,2n;8n\rangle_{\rm GS,1/2}$, cf
Eq. (\ref{4states}), correspond to quantum numbers 
\bea
|2n,2n;8n\rangle_{\rm GS,1}
:&& m_\uparrow=1=m'_\downarrow\ ,\quad
m_\downarrow=m'_\uparrow=0,\quad \Leftrightarrow \quad 
q_{\up,1}=\bar{q}_{\down,1}=\frac{\sqrt{\pi}}{2}\ ,\
q_{\down,0}=\bar{q}_{\up,0}=-\frac{\sqrt{\pi}}{2}\ ,
\nn
|2n,2n;8n\rangle_{\rm GS,2}
:&& m_\downarrow=1=m'_\uparrow\ ,\quad m'_\downarrow=m_\uparrow=0, 
\quad \Leftrightarrow \quad q_{\down,1}=\bar{q}_{\up,1}=\frac{\sqrt{\pi}}{2}\ ,\
q_{\up,0}=\bar{q}_{\down,0}=-\frac{\sqrt{\pi}}{2}.
\label{12bos}
\eea
The calculation of correlators in any of the sixteen ground states for
$L=8n$ proceeds in the same way as for $L=8n+4$, cf
Sec. \ref{Cor8n4}. The only difference arises from factors like
\be
\langle q_{\sigma,m},\bar{q}_{\sigma,m'}|e^{-i\sqrt{4\pi}Q_\sigma
  {x}/{\LL}} 
|q_{\sigma,m},\bar{q}_{\sigma,m'}\rangle\ .
\ee
If we consider the state with $m=m'=0$ then this produces an
additional factor $e^{i\pi x/\LL}$ for the two point function of right moving fermions
and a factor $e^{-i\pi x/\LL}$ for the left movers. We then obtain the result
\bea
{}_{\rm GS}\langle 2n-1,2n-1;8n|c^\dagger_{j+\ell,\sigma}
c_{j,\sigma}|2n-1,2n-1;8n\rangle_{\rm GS}&\sim& 
\frac{\sin(\tilde{k}_F \ell)}{L\sin(\pi \ell/L)},
\eea
where $\tilde{k}_F$ is given by \fr{ktilde}. Hence we again reproduce the
correct lattice result.
\subsubsection{Excited State for $L=8n+4$}
The excited state $|{\rm EXC}\rangle$ in Eq. (\ref{exc1}) corresponds
in the field theory limit to a linear combination of states with 
\be
\Delta N_\up=\Delta N_\down=1\ ,\quad
D_\up=-D_\down=\frac{1}{2}\ , \quad \Leftrightarrow \quad
m_\up=1\ ,\ m'_\up=0\ ,\quad
m_\down=0\ ,\ m'_\down=1\ ,
\ee
and
\be
\Delta N_\up=\Delta N_\down=1\ ,\quad
D_\up=-D_\down=-\frac{1}{2}\ ,\quad \Leftrightarrow \quad
m_\up=0\ ,\ m'_\up=1\ ,\quad
m_\down=1\ ,\ m'_\down=0\ .
\ee
In order to obtain a representation on the bosonic Fock space, it is
convenient to employ mode the expansions in Euclidean space
\bea
\varphi_\sigma(z)&=&\varphi_{\sigma,0}+\frac{i}{2\pi}\ln(z)Q_\sigma
+\sum_{n=1}^\infty\frac{1}{\sqrt{4\pi n}}\left[
z^{-n}a_{\sigma,R,n}+z^na^\dagger_{\sigma,R,n}\right] ,\nn
\bar{\varphi}_\sigma(\bar{z})&=&\bar{\varphi}_{\sigma,0}-\frac{i}{2\pi}
\ln(\bar{z})\bar{Q}_\sigma+\sum_{n=1}^\infty\frac{1}{\sqrt{4\pi n}}\left[  
\bar{z}^{-n}a_{\sigma,L,n}+\bar{z}^{n}a^\dagger_{\sigma,L,n}\right] ,
\eea
where we have defined complex coordinates
\be
z=e^{\frac{2\pi}{\LL} (v\tau-ix)}\ ,\quad
\bar{z}=e^{\frac{2\pi}{\LL} (v\tau+ix)}\ .
\ee
Let us now consider the particular class of states
\be
|\alpha,\bar{\alpha},\sigma\rangle
\equiv\lim_{z,\bar{z}\to 0}e^{i\alpha\varphi_\sigma(z)+
i\bar{\alpha}\bar{\varphi}_\sigma(\bar{z})}|0\rangle.
\label{rad1}
\ee
In the limit $z,\bar{z}\to 0$ the oscillator modes drop out as only
the annihilation operators survive. Thus
\bea
Q_\sigma|\alpha,\bar{\alpha},\sigma\rangle&=&
-\frac{\alpha}{2}
|\alpha,\bar{\alpha},\sigma\rangle\ ,\nn
\bar{Q}_\sigma|\alpha,\bar{\alpha},\sigma\rangle&=&
\frac{\bar{\alpha}}{2}
|\alpha,\bar{\alpha},\sigma\rangle\ ,\nn
a_{\sigma,R/L,n}|\alpha,\bar{\alpha},\sigma\rangle&=&0.
\label{rad2}
\eea
Using these results, we conclude that in the bosonized theory 
the excited state $|{\rm EXC}\rangle$ in Eq. (\ref{exc1}) corresponds to
\be
ae^{-i\sqrt{4\pi}\varphi_\up(0)+i\sqrt{4\pi}\bar{\varphi}_\down(0)}|0\rangle
+be^{-i\sqrt{4\pi}\varphi_\down(0)+i\sqrt{4\pi}\bar{\varphi}_\up(0)}|0\rangle\ ,
\ee
where $a$ and $b$ are complex numbers. As our state should be
symmetric under interchange of up and down spins we must have that $a=b$.
In terms of spin and charge bosons we conclude that
\be
|{\rm EXC}\rangle\sim
e^{-i\sqrt{2\pi}\Theta_c(0,0)}\ \cos\left(\sqrt{2\pi}\Phi_s(0,0)\right)|0\rangle.
\ee
\subsubsection{Ground State $|+\rangle$ for $L=8n$}
Finally, we are in a position to express our particular state of
interest $|+\rangle$ \fr{desired} in the bosonized theory.
Employing Eqns \fr{rad1} and \fr{rad2}, we conclude that the states
\fr{12bos} correspond to
\bea
|2n,2n;8n\rangle_{\rm GS,1}&\sim& e^{-i\sqrt{\pi}\big(\varphi_\up(0,0)-\varphi_\down(0,0)
-\bar{\varphi}_\down(0,0)+\bar{\varphi}_\up(0,0)\big)}|0\rangle
\sim e^{-i\sqrt{2\pi}\Phi_s(0,0)}|0\rangle\ ,\nn
|2n,2n;8n\rangle_{\rm GS,2}&\sim& e^{-i\sqrt{\pi}\big(\varphi_\down(0,0)-\varphi_\up(0,0)
-\bar{\varphi}_\up(0,0)+\bar{\varphi}_\down(0,0)\big)}|0\rangle\sim
e^{i\sqrt{2\pi}\Phi_s(0,0)}|0\rangle\ ,
\label{rad3}
\eea
where $|0\rangle$ is the boson vacuum (which has zero mode eigenvalues
zero). We note that for $L=8n$ the vacuum state $|0\rangle$ is not an allowed
state of the compact boson theory, and expressions \fr{rad3} must be
understood on an extended bosonic Fock space.
We conclude that the state \fr{desired} has the bosonic representation 
\be
|+\rangle\sim
\cos\left(\sqrt{2\pi}\Phi_s(0,0)\right)|0\rangle.
\label{EXC8n}
\ee

\section{Hubbard Model}
\label{sec:hm}
We now turn to the Hubbard model
\be
H(U)=-t\sum_{j,\sigma}(c^\dagger_{j,\sigma}c_{j+1,\sigma}+
c^\dagger_{j+1,\sigma}c_{j,\sigma})+U\sum_j n_{j,\uparrow}\ n_{j,\downarrow}-\mu\sum_j n_j\ ,
\label{HHubb2}
\ee
where $n_{j,\sigma}=c^\dagger_{j,\sigma}c_{j,\sigma}$ and
$n_j=n_{j,\uparrow}+n_{j,\downarrow}$ and we impose periodic boundary
conditions. The Hubbard model is solvable by the Bethe Ansatz, and
its particular eigenstates relevant to our discussion are parametrized
in terms of the solutions $\{\Lambda_\alpha,k_j \}$ of the following
set of coupled  Bethe Ansatz Equations equations \cite{lw,book} 
\begin{eqnarray} 
&&     k_j L  =  2 \pi I_j - \sum_{\alpha = 1}^{M}
                 \theta \left(
		 \frac{\sin k_j - \Lambda_\alpha}{u} \right),\quad
               j=1,\ldots,N\ ,
		 \nn
&&     \sum_{j=1}^{N} \theta \left(
		 \frac{\Lambda_\alpha - \sin k_j}{u} \right)  = 
		 2 \pi J_\alpha+
		 \sum_{\beta = 1}^{M}
		 \theta \left(
		 \frac{\Lambda_\alpha - \Lambda_\beta}{2u} \right), \, \alpha=1,\ldots,M.
\label{BAE}
\end{eqnarray}
Here $u={U}/{4t}$ is a dimensionless interaction strength, the length
of the lattice $L$ is taken to be even, $\theta(x) =2 \arctan(x)$ and
$N=N_\up+N_\down$ . 
The quantum numbers $I_j$, ${J}_\alpha$ are integer or
half-odd integer numbers that arise due to the multivaluedness of the
logarithm. They are subject to the ``selection rules''
\be
I_j\ {\rm is}\ \bigg\{\begin{array}{l l}
{\rm integer} &\text{if}\ M\ \text{is even}\\
{\rm half-odd\ integer} &\text{if}\ M\ \text{is odd},\\
\end{array}
\label{int-hoi1}
\end{equation}
\be
J_\alpha\ {\rm is}\ \bigg\{\begin{array}{l l}
{\rm integer} &\text{if}\ N-M\ \text{is odd}\\
{\rm half-odd\ integer} &\text{if}\ N-M\ \text{is even},\\
\end{array}
\label{int-hoi2}
\end{equation}
and have ranges
\be
-\frac{L}{2}<I_j\leq\frac{L}{2}\ ,\qquad
|J_\alpha|\leq \frac{1}{2}(N-M-1)\ .
\label{ranges}
\ee
The energy and momentum of such Bethe ansatz states are
\bea
E&=&-\sum_{j=1}^N\left[2t\cos(k_j)+\mu\right]\ ,\nn
P&=&\sum_{j=1}^N k_j\equiv\frac{2\pi}{L}\left[
\sum_{j=1}^NI_j+\sum_{\alpha=1}^MJ_\alpha\right].
\label{EP}
\eea
Following Refs. \cite{hwt}, we define \emph{regular} Bethe
Ansatz states as eigenstates of $H(U)$ arising from solutions of
\fr{BAE} with $2M\leq N$, where all $k_j$ and $\Lambda_\alpha$ are
finite. We denote these states by
\be
|\{I_j\};\{J_\alpha\}\rangle_{\rm reg}.
\ee
It was shown in Refs. \cite{hwt} that regular Bethe Ansatz
states are highest weight states with respect to the SO(4) symmetry
\cite{so4} of the Hubbard model, i.e.
\bea
\eta|\{I_j\};\{J_\alpha\}\rangle_{\rm reg}&=&0\ ,\nn
S^+|\{I_j\};\{J_\alpha\}\rangle_{\rm reg}&=&0\ ,
\eea
where
\be
S^+=\sum_{j=1}^Lc^\dagger_{j,\uparrow}c_{j,\downarrow}\ ,\quad
\eta=\sum_{j=1}^L(-1)^jc_{j,\uparrow}c_{j,\downarrow}\ .
\ee
A complete set of eigenstates is obtained by acting with lowering
operators on the lowest weight states \cite{hwt}
\be
\big(S^-\big)^m\big(\eta^\dagger\big)^n|\{I_j\};\{J_\alpha\}\rangle_{\rm 
  reg}. 
\ee
\subsection{Quarter-Filled Ground State for $L=4\ {\rm mod}\ 8$.}
For $L=8n+4$ we have $N={L}/{2}=4n+2$ and $M=2n+1$. Hence the
$I_j$'s are half-odd integers and the $J_\alpha$'s are integers. The
ground state is characterised by the quantum numbers
\bea
I_j&=&-2n-\frac{3}{2}+j\ ,\ j=1,\ldots,4n+2\ ,\nn
J_\alpha&=&-n-1+\alpha\ ,\ \alpha=1,\ldots,2n+1.
\eea
Importantly, the distributions of $I_j$ and $J_\alpha$ are
\emph{symmetric} around zero. The ground state is a regular Bethe
Ansatz state and has total spin $S=0$, i.e.
\be
|{\rm GS}\rangle=|\{I_1,\ldots,I_{\frac{L}{2}}\};\{J_1,\ldots,
J_{\frac{L}{4}}\}\rangle_{\rm reg}.
\ee
The excitation spectrum relative to the ground state has been derived
in Ref.~\cite{FSspectrum} and is given by
\be
\Delta E=\frac{2\pi  v_c}{L}
\left[\frac{(\Delta N_c)^2}{4\xi^2}
+\xi^2\Big(D_c+\frac{D_s}{2}\Big)^2+N_c^++N_c^-\right]
+\frac{2\pi  v_s}{L}
\left[\frac{\left(\Delta {N}_s-\frac{\Delta {N}_c}{2}\right)^2}{2}
+\frac{D_s^2}{2}+N_s^++N_s^-\right]+o(L^{-1}), 
\label{ECFT}
\ee
where $\Delta N_\a$, $2D_\a$ and $N_\a^\pm$ are integer ``quantum
numbers'' subject to the selection rules 
\be
N_\a^\pm \in \mathbb{N}_0\ ,\quad \Delta N_\a\in \mathbb{Z}\ ,\quad
D_c=\frac{\Delta N_c+\Delta N_s}{2}\text{mod}\ 1\ ,\
D_s=\frac{\Delta N_c}{2}\text{mod}\ 1.
\ee
Here $\xi=\xi(Q)$ is obtained from the solution of the integral equation
\be
\xi(k)=1+\int_{-Q}^Qdk'\ \cos(k')\ R(\sin(k)-\sin(k'))\ \xi(k'),
\label{xiB0}
\ee
where
\be
R(x)=\int_{-\infty}^\infty \frac{d\omega}{2\pi}\frac{e^{i\omega x}}
{1+\exp(2u|\omega|)}.
\label{rofx}
\ee

\subsection{Quarter-Filled Ground State for $L=0\ {\rm mod}\ 8$.}
For $L=8n$ we have $N={L}/{2}=4n$ and $M=2n$. Hence the $I_j$'s are
integers and the $J_\alpha$'s are half-odd integers. One may naively
expect the ground state to be obtained by choosing either
\bea
I_j^{(1)}&=&-2n+j\ ,\ j=1,\ldots,4n\ ,\nn
J_\alpha^{(1)}&=&-n-\frac{1}{2}+\alpha\ ,\ \alpha=1,\ldots,2n,
\label{state1}
\eea
or
\bea
I_j^{(2)}&=&-2n-1+j\ ,\ j=1,\ldots,4n\ ,\nn
J^{(2)}_\alpha&=&-n-\frac{1}{2}+\alpha\ ,\ \alpha=1,\ldots,2n.
\label{state2}
\eea
Comparison of their energies to numerical results for the ground state
energy shows that this is not the case. The ground state is in
fact obtained as follows. We start by considering regular Bethe Ansatz
states with one fewer down spin, i.e. $N={L}/{2}=4n$ and $M=2n-1$. Now
the $I_j$'s are half-odd integers and the $J_\alpha$'s are integers. The 
lowest energy regular Bethe Ansatz state corresponds to the choice
\bea
I_j^{(0)}&=&-2n-\frac{1}{2}+j\ ,\ j=1,\ldots,4n\ ,\nn
J^{(0)}_\alpha&=&-n+\alpha\ ,\ \alpha=1,\ldots,2n-1.
\label{state0}
\eea
We denote its energy by $E\big(\{I_j^{(0)}\};\{J_\alpha^{(0)}\}\big) $.
The corresponding state is an eigenstate of $S^z$ with eigenvalue $1$
\be
S^z|\{I_j^{(0)}\};\{J_\alpha^{(0)}\}\rangle_{\rm reg}=
|\{I_j^{(0)}\};\{J_\alpha^{(0)}\}\rangle_{\rm reg}.
\ee
As we are dealing with a regular Bethe Ansatz state, the theorem of
Ref.~\cite{hwt} applies and we may conclude that we are dealing with the
highest weight state of a spin-SU(2) triplet. The ground state of the
quarter-filled Hubbard model with $L=8n$ is then obtained as the
$S^z=0$ state of this spin triplet
\be
|{\rm GS}\rangle=S^-|\{I_j^{(0)}\};\{J_\alpha^{(0)}\}\rangle_{\rm reg}.
\label{GS8n}
\ee

\section{Luttinger Liquid Description of the Hubbard Chain}
\label{sec:LL}

The bosonization of the Hubbard model proceeds by first bosonizing the
theory at $U=0$, and then taking the interactions into account \cite{affleck}.
The result of this analysis is a spin-charge separated Luttinger
liquid Hamiltonian of the form
\be
{\cal H}=\sum_{\alpha=c,s}\frac{v_\alpha}{2}\int dx \left[(\partial_x\Phi_\alpha)^2+ (\partial_x\Theta_\alpha)^2\right],
\label{Hboson}
\ee
where the spin and charge fields are now given by
\bea
\Phi_c&=&\frac{\Phi_\uparrow+\Phi_\downarrow}{\sqrt{2K}}\ ,\quad
\Theta_c=\sqrt{\frac{K}{2}}
 (\Theta_\uparrow+\Theta_\downarrow)\ ,\nn
\Phi_s&=&\frac{\Phi_\uparrow-\Phi_\downarrow}{\sqrt{2}}\ ,\quad
\Theta_s=\frac{\Theta_\uparrow-\Theta_\downarrow}{\sqrt{2}}\ .
\eea
The normalization of $H$ is such that Eq. (\ref{norm1}) is satisfied
by the fields $\Phi_{c,s}$. The Luttinger parameter $K$ and the
velocities $v_{c,s}$ depend on the interaction strength $U$ (actually
on $U/t$) and they can be calculated by solving appropriate integral
equations  (see e.g. \cite{book}). In fact $K$ is related to the
solution $\xi(k)$ of the integral equation (\ref{xiB0}) by
\be
K=\frac{\xi(Q)^2}{2}.
\ee
The mode expansions for the spin and charge bosons are
\bea
\varphi_{\tt a}(x,t)&=&\varphi_{\tt a,0}+\frac{x-vt}{\LL}Q_{\tt a}
+\sum_{n=1}^\infty\frac{1}{\sqrt{4\pi n}}\left[
e^{i\frac{2\pi n}{\LL}(x-vt)}a_{{\tt a},R,n}+e^{-i\frac{2\pi
    n}{\LL}(x-vt)}a^\dagger_{{\tt a},R,n}\right] ,\nn
\bar{\varphi}_{\tt a}(x,t)&=&\bar{\varphi}_{{\tt a},0}+\frac{x+vt}{\LL}
\bar{Q}_{\tt a}+\sum_{n=1}^\infty\frac{1}{\sqrt{4\pi n}}\left[  
e^{-i\frac{2\pi n}{\LL}(x+vt)}a_{{\tt a},L,n}+e^{i\frac{2\pi
    n}{\LL}(x+vt)}a^\dagger_{{\tt a},L,n}\right] ,
\eea
where ${\tt a}=c,s$ and the zero momentum mode operators have
commutation relations
\be
[\varphi_{{\tt a},0},Q_{\tt a}]=-\frac{i}{2}=
-[\bar{\varphi}_{{\tt a},0},\bar{Q}_{\tt a}].
\ee
The spin and charge zero mode operators are related to the up and down
zero mode operators by the canonical transformation
\bea
Q_c&=&\frac{K+1}{\sqrt{4K}}
\frac{Q_\up+Q_\down}{\sqrt{2}}+
\frac{1-K}{\sqrt{4K}}
\frac{\bar{Q}_\up+\bar{Q}_\down}{\sqrt{2}}\ ,\\
\bar{Q}_c&=&\frac{1-K}{\sqrt{4K}}
\frac{Q_\up+Q_\down}{\sqrt{2}}+
\frac{K+1}{\sqrt{4K}}
\frac{\bar{Q}_\up+\bar{Q}_\down}{\sqrt{2}}\ ,\\
Q_s&=&\frac{Q_\up-Q_\down}{\sqrt{2}}\ ,\\
\bar{Q}_s&=&
\frac{\bar{Q}_\up-\bar{Q}_\down}{\sqrt{2}}\ .
\eea
The mode expansion of the Hamiltonian is
\bea
{\cal H}&=&\sum_{{\tt a}=c,s}\frac{v_{\tt a}}{\LL}\left[
Q_{\tt a}^2+\bar{Q}_{\tt a}^2+
\sum_{n>0}2\pi n\left(a^\dagger_{{\tt a},R,n}a_{{\tt a},R,n}
+a^\dagger_{{\tt a},L,n}a_{{\tt a},L,n}\right)\right].
\label{Hmodes}
\eea
Imposing periodic boundary conditions on the lattice fermions leads to
the quantization of the zero mode eigenvalues in the same way as in
the $U=0$ case. In particular we again have
\be
e^{i\sqrt{4\pi}Q_\sigma}
=e^{-i\sqrt{4\pi}\bar{Q}_\sigma}=
\begin{cases}
1 &\text{if } L=8n+4\ ,\\
-1 &\text{if } L=8n.
\end{cases}
\ee
The corresponding eigenvalues are the same as for $U=0$, see eqns
\fr{quant1}, \fr{quant2}.
\subsection{Finite-size spectrum for $L=8n+4$}
Using the quantization conditions for the zero mode operators in the
mode expansion \fr{Hmodes}, we obtain a finite-size spectrum of the form
\bea
E&=&\frac{2\pi v_c}{\LL}\left[\frac{1}{8K}
(m_\up+m_\down+m'_\up+m'_\down)^2
+{\frac{K}{8}}
(m_\up+m_\down-m'_\up-m'_\down)^2
+\sum_{n>0}n\left[M_{n,c}^++M_{n,c}^-\right]
\right]\nn
&+&\frac{2\pi v_s}{\LL}\left[
\left(\frac{m_\up-m_\down}{2}\right)^2
+\left(\frac{m'_\up-m'_\down}{2}\right)^2
+\sum_{n>0}n\left[M_{n,s}^++M_{n,s}^-\right]\right].
\label{ELL}
\eea
Defining new quantum numbers
\be
\Delta N_c=m_\uparrow+m_\downarrow+m'_\uparrow+m'_\downarrow\ ,\quad
\Delta N_s=m_\downarrow+m'_\downarrow\ ,\quad
2D_c=m_\downarrow-m'_\downarrow\ ,\quad
2D_s=m_\uparrow-m_\downarrow-m'_\uparrow+m'_\downarrow\ ,
\ee
and using that $K=\xi^2/2$, we recover the expression
\fr{ECFT} (with the correct selection rules for $D_{c,s}$) obtained
directly from the Bethe Ansatz. 

\subsection{Finite-size spectrum for $L=8n$}
Here we can proceed analogously, the only difference being that we the
eigenvalues are shifted
\be
m_\sigma\rightarrow m_\sigma-\frac{1}{2}\ ,\quad
m'_\sigma\rightarrow m'_\sigma-\frac{1}{2}.
\ee
The finite-size spectrum is then obtained by carrying out these
substitions in Eqn \fr{ELL}. Analysis of the resulting energy levels
shows that there are \emph{four} ground states with $E=\pi v_s/\LL$
characterized by quantum numbers
\bea
m_\up&=&m'_\up=1\ ,\ m_\down=m'_\down=0\ ,\nn
m_\up&=&m'_\up=0\ ,\ m_\down=m'_\down=1\ ,\nn
m_\sigma&=&m'_{-\sigma}=1\ ,\ m_{-\sigma}=m'_\sigma=0\ .
\label{1and2}
\eea
All four states occur in the sector $N_\uparrow+N_\downarrow=L/2$, but
they differ in their $S^z$ eigenvalues. In particular, the last two
states both have $S^z=0$.
\subsection{Ground State for $L=8n$}
We know from our analysis of the Hubbard model, that the unique ground
state occurs in the sector with $S^z=0$, and is part of a spin triplet.
To order ${\cal O}(L^{-1})$ this state is degenerate with a spin
singlet, which explains the four-fold degeneracy observed in the
Littinger liquid description. As we are ultimately interested in the
ground state of the Hubbard model, we require an appropriate
linear combination of the two states \fr{1and2}. Recalling that the
zero mode eigenvalues are  
\bea
q_\up=-q_\down=\frac{\sqrt{\pi}}{2}\ ,\quad
\bar{q}_\down=-\bar{q}_\up=\frac{\sqrt{\pi}}{2}\quad\text{and}\quad
q_\up=-q_\down=-\frac{\sqrt{\pi}}{2}\ ,\quad
\bar{q}_\down=-\bar{q}_\up=-\frac{\sqrt{\pi}}{2},
\eea
we may use \fr{rad1}, \fr{rad2} together with the requirement that the
state must be symmetric under the interchange of up and down spins to
conclude that the Hubbard model ground state for $L=8n$ corresponds to
\be
|{\rm GS}\rangle\sim
\cos\big(\sqrt{2\pi}\Phi_s(0,0)\big)|0\rangle.
\label{gs2a}
\ee

\subsection{Excited States for $L=8n+4$}
For the tight-binding model we showed that correlators calculated in
the particular excited states \fr{exc1} for lattice lengths $L=8n+4$
are essentially the same (apart from a shift in $k_F$) as ground state
correlators for $L=8n$. This suggests that for the Hubbard model there
might be excited states on $L=8n+4$ site lattices, whose
EEs exhibit the same kind of additional
contribution as the ground state of the $L=8n$ site system. In order
to investigate this idea, we consider the degenerate excited states
characterized by the quantum numbers
\bea
\Delta N_\up=\Delta N_\down=1\ ,\quad
D_\up=-D_\down=\frac{1}{2}\ ,&\Leftrightarrow&
m_\up=1\ ,\ m'_\up=0\ ,\quad
m_\down=0\ ,\ m'_\down=1\ ,\nn
\Delta N_\up=\Delta N_\down=1\ ,\quad
D_\up=-D_\down=-\frac{1}{2}\ ,&\Leftrightarrow&
m_\up=0\ ,\ m'_\up=1\ ,\quad
m_\down=1\ ,\ m'_\down=0\ .
\eea
Both states are in the $S^z=0$ sector with two particles added
relative to the ground state. The appropriate linear combinations that
correspond to eigenstates of the total spin ${\bf S}^2$, i.e. spin
triplet and singlet states, are
\bea
|{\rm triplet}\rangle&\sim&
e^{-i\sqrt{2\pi/ K}\Theta_c(0,0)}\ \cos\left(\sqrt{2\pi}\Phi_s(0,0)\right)|0\rangle\ ,\nn
|{\rm singlet}\rangle&\sim&
e^{-i\sqrt{2\pi/ K}\Theta_c(0,0)}\ \sin\left(\sqrt{2\pi}\Phi_s(0,0)\right)|0\rangle\ .
\label{exc2}
\eea
In the following we will determine the entanglement entropies in these
two states and compare them to their corresponding ground state values.

\subsection{Marginally irrelevant perturbation and ``$L$-dependent
  exponents''} 
\label{ssec:Keff}
In the bosonization approach reviewed above the Luttinger parameter in
the spin sector is fixed by the $SU(2)$ symmetry to be $K_s=1$. It is
however well known, that the spin sector is affected by the presence
of a marginally irrelevant perturbation\cite{affleck}, which gives rise
to multiplicative logarithmic corrections in correlation functions of
local operators \cite{logs}.
A consequence of these corrections is that numerical results obtained
for finite-size systems are generally not well described by the
power-laws extracted from Luttinger liquid theory. This can be
understood by employing renormalization group methods: the marginally irrelevant
interaction essentially gives rise to a scale dependence of the spin Luttinger
parameter. A proper treatment involves solving the appropriate
Callan-Symanzik equation for the particular quantity of interest. This
is beyond the scope of our work, and we instead resort to
semi-phenomenological considerations. We take the lattice length $L$
to be our RG scale, and are interested in the regime $32\leq L\leq
104$ relevant for our numerical studies of entanglement entropies.
The idea is then to fit the finite-size ground state energy for $L=8n$
to the form (see also Appendix \ref{Appa}) 
\be
\frac{E}L=\alpha+\beta L^{-2}.
\label{Efit}
\ee
For asymptotically large $L$ (so that the logarithmic corrections are
negligible) the coefficient $\beta$ is related to the Luttinger
parameter of the spin sector by
\be
K_s=\big(\pi v_s\big)^{-1}
\left(\beta+\frac{\pi(v_s+v_c)}{6}\right)=1.
\ee
In presence of the marginal perturbation, the coefficient $\beta$
extracted from \fr{Efit} becomes $L$ and $U$ dependent and we then can 
define an effective spin Luttinger parameter $K_s^{\rm eff}(U)$ by 
\be
K^{\rm eff}_s(U)=\big(\pi  v_s\big)^{-1}\left(\beta+\frac{\pi(v_s+v_c)}{6}\right).
\ee
In the limit $L\to\infty$ we must have $K_s^{\rm eff}(U)=1$ by $SU(2)$
symmetry. By analysing the ground state energy for a quarter-filled
Hubbard chain for lattice lengths between $L=32$ and $L=104$ (which
are the typical lengths considered in the numerical simulations in
the following sections), we obtain the following results 
\be
\begin{tabular}{l|l|l|l|l|l}
\hline
U & 0.5 & 1&2 & 4 & 16 \\
\hline
$K_s^{\rm eff}(U)$ &0.958 & 0.933& 0.906 & 0.887 & 0.873\\
\hline
\end{tabular}
\label{table:Ks}
\ee
We note that, as expected, the effect of the marginally irrelevant
corrections increases substantially with the Hubbard coupling $U$.

The effects of the marginally irrelevant perturbation on the
entanglement entropies for finite chains with $32\leq L\leq 104$ can
then be estimated by replacing $K_s=1$ with $K_s^{\rm eff}(U)$. In
particular, the calculation of the ground state entanglement would be
modified by replacing the Luttinger liquid representation \fr{gs2a} of
the ground state by
\be
\cos\left(\sqrt{2\pi{K^{\rm eff}_s(U)}}\Phi_s(0,0)\right)|0\rangle\ .
\label{Omod}
\ee
Analogous replacement would be done for the excited states \fr{exc2}.
\section{CFT Approach to Entanglement Entropies}
\label{sec:CFT}

In the previous sections we have derived explicit representations of
the ground state (and some low-lying excited states) of the
quarter-filled Hubbard model in terms of the underlying bosonic CFT. 
The results are summarized as follows:
\begin{enumerate}
\item{} Ground state for $L=8n+4$
\be
|{\rm GS}\rangle\sim |0\rangle.
\ee
\item{} Ground state for $L=8n$
\be
|{\rm GS}\rangle\sim
\cos\big(\sqrt{2\pi}\Phi_s(0,0)\big)|0\rangle\equiv {\cal O}_1|0\rangle.
\label{gs4}
\ee
\item{} Excited states for $L=8n+4$
\bea
|{\rm triplet}\rangle&\sim&
e^{-i\sqrt{2\pi K}\Theta_c(0,0)}\ \cos\big(\sqrt{2\pi}\Phi_s(0,0)\big)|0\rangle 
\equiv {\cal O}_2|0\rangle ,\nn
|{\rm singlet}\rangle&\sim&
e^{-i\sqrt{2\pi K}\Theta_c(0,0)}\ \sin\big(\sqrt{2\pi}\Phi_s(0,0)\big)|0\rangle
\equiv {\cal O}_3|0\rangle.
\label{excgood}
\eea
\end{enumerate}
The next step is to calculate entanglement properties within the CFT framework.
In the first case (ground state for $L=8n+4$) the entanglement
entropies follow from the general CFT vacuum result
\fr{criticalent}. In all other cases one is dealing with entanglement
entropies of particular excited states (in the second case the state
corresponds to an excitation on an extended Hilbert space, as
discussed above). This observation allows us to make use of results
for entanglement entropies in low-lying excited states in CFTs. 
A general approach to the the latter problem has been developed
by Alcaraz et al. \cite{abs-11,abs-12} (see also \cite{exc-other,exc-other2,exc-other3,exc-other4,exc-other5,exc-other6,exc-other7} for
other studies of the entanglement entropies in excited states of many
body systems) and their main result can be summarized as follows. 
The n'th R\'enyi entropy for an excited state of the form ${\cal
  O}(0,0)|0\rangle$ is given by 
\be
S_n=\frac{c}{6}\Big(1+\frac{1}{n}\Big)\ln\left[\frac{L}{\pi}
\sin\left(\frac{\pi\ell}{L}\right)\right] +\frac{1}{1-n}\ln\left[F_n(\ell/L)\right]+c_n +o(L^0),
\label{SnCFT}
\ee
where $c_n$ are ${\cal O}$-independent constants (which are the same
as for the EEs in the CFT vacuum), and the scaling functions $F_n^{\cal
  O}(x)$ are given by 
\be
F_{n}^{\cal O}(x)=\frac{\left\langle\displaystyle\prod_{k=0}^{n-1}{\cal O}\Big(\frac{\pi}{n}(x+2k)\Big)
{\cal O}^\dagger\Big(\frac{\pi}{n}(-x+2k)\Big)\right\rangle}
{n^{2n(h+\bar{h})}\langle{\cal O}(\pi x){\cal O}^\dagger(-\pi x)\rangle^n}\ . 
\label{FnO}
\ee
Here $h$ and $\bar{h}$ are the conformal dimensions of the operator
${\cal O}$. For the Hubbard model, we have $c=2$ and are interested in
the operators ${\cal O}_j(x)$ in Eqn \fr{excgood}. These operators
factorize into a spin and a charge part, which in turn leads to the
factorization of the scaling function $F^{\cal O}_n(x)$. Importantly,
according to a result obtained by Alcaraz et al. \cite{abs-11} the
scaling function for vertex operators is trivial
\be
F^{e^{i{\alpha\Theta_c}}}_{n}(x)=1.
\ee
In our case this implies that the charge sector does not contribute to
the scaling function. Furthermore, the structure of the expectation
value of \fr{FnO} is such that the results for ${\cal
  O}=\cos(\alpha\Phi_s)$ and ${\cal O}=\sin(\alpha\Phi_s)$ are identical.
Combining these observations we conclude that \emph{the extra
contribution in \fr{SnCFT} is identical for the $L=8n$ ground state
(case 2) and the $L=8n+4$ excitations (case 3) considered above}. This
provides a first prediction for Hubbard model EEs.

\subsection{CFT calculation of the R\'enyi entropies}

In order to evaluate the correlation function appearing in the scaling
function \fr{FnO} we can use the standard Coulomb gas identity
\be
\langle\prod_{j=1}^{2n}:2\cos\big(\beta\Phi_s(x_j)\big):\rangle=
\sum_{\sigma_1,\ldots,\sigma_{2n}}
\delta_{\sigma_1+\sigma_2+\dots+\sigma_{2n},0}\prod_{i<j}\left
|2\sin\big(\frac{x_i-x_j}{2}\big)\right|^{\sigma_i\sigma_j \frac{\beta^2}{2\pi}},
\label{Cgas}
\ee
where the $x_j$'s can be read from Eq. \fr{FnO} and they explicitly are
\be
x_{2j-1}=\frac{\pi x}{n} + \frac{2\pi(j - 1)}{n}\ ,\quad
x_{2j}=-\frac{\pi x}{n} + \frac{2\pi(j - 1)}{n}\ ,\quad
j=1,\ldots,n.
\ee

Since in Eqs. \fr{gs4} and \fr{excgood} we have $\beta^2=2\pi$,
the function $F_n(x)$ can be readily obtained for low values of $n$ by 
direct computation, giving, up to $n=6$:
\bea
F_2(x)&=&
\frac18[7 + \cos(2 \pi x)],\nn
F_3(x)&=& 
\frac19[7+2\cos(2\pi x)] ,\nn
F_4(x)&=&
\frac{1}{2048}[1435+604\cos(2\pi x)+9\cos(4\pi x)]\ ,\nn
F_5(x)&=& \frac1{625}[399+218\cos(2\pi x)+8\cos(4\pi x)]\ ,\nn
F_6(x)&=&
\frac{1}{165888}[ 97430 + 64439 \cos(2\pi x) + 3994 \cos(4\pi x) +  25 \cos(6\pi x)]\ ,
\label{F2-F6}
\eea
where we introduced 
\be
s_n=\sin\big(\frac{\pi x}{n}\big).
\ee
It is worth mentioning that the expansion of $F_n(x)$ for small $x$ is
\be
F_n(x)=1+\frac16\Big( n-\frac1n\Big) (\pi x)^2+ O(x^4),
\label{smallx}
\ee
in agreement with the general result reported in Ref. \cite{abs-11}.
We observe that our expressions for $F_n(x)$ for $1\leq n\leq 6$ are
equal to the square root of the scaling function $F_n^\Upsilon(x)$ for
the operator $\Upsilon=i\partial \phi$ in a compactified boson theory,
cf Eq. (56) in Ref.~\cite{abs-12}. Although we are unsure whether
there is a deep connection between the two, we can safely conjecture
that this relationship holds for arbitrary $n$, i.e.
\be
[F_n(x)]^2= F_n^\Upsilon(x)\,.
\ee
In Ref. \cite{abs-12} a determinant representation for the function
$F_n^\Upsilon(x)$ has been obtained
\be
F_{n}^\Upsilon(x)=  \left(\frac1n\sin{(\pi x)}\right)^{2n}\det {\mathbb H}\,.
\label{predabs}
\ee
where $ {\mathbb H}$ is a $2 n\times 2 n$ matrix with elements 
\be
{\mathbb H}_{jk}=\begin{cases} \frac1{\sin[(z_j-z_k)/2]}& {\rm if}\; j\neq k\\ 0&{\rm if}\; j= k\end{cases}
\quad {\rm and} \quad 
z_j=\begin{cases}\pi(2j-2+x)/n& {\rm if}\; j\leq n\\ \pi(2j-2-x)/n& {\rm if}\; j> n\end{cases}\,.
\label{Hmat}
\ee
This representation holds only for integer values of $n$, but in the next
subsection we will provide its analytic continuation to arbitrary $n$.

\subsection{The analytic continuation and the von Neumann Entropy}

In order to find the analytic continuation of $F_n(x)$ to arbitrary $n$, let us start by 
re-organising the order of row and column indices of the matrix ${\mathbb H}$ in Eq. \fr{Hmat}
rewriting it in the block form
\be
\mathbb H= \left(
\begin{array}{cc}
\mathbb A& \mathbb B\\
-\mathbb B^T& \mathbb A
\end{array}
\right),
\ee
where the matrix elements are ($i,j=1,\ldots,n$)
\be
\mathbb A_{ij}=\begin{cases}
0 &\text{if } i=j,\\
\frac1{\sin[\pi(j-i)/n ]} & \text{else}\ ,
\end{cases}\ , \qquad
\mathbb B_{ij}= \frac1{\sin[\pi(j-i-x)/n ]}\ .
\ee 
 It is straightforward to see that $\mathbb A$ and $\mathbb B$ commute and so
 \be
 \det \mathbb H=\det[\mathbb A^2+\mathbb B^T \mathbb B].
 \ee
A direct calculation shows also that
 \be
 \mathbb B^T \mathbb B=\a \mathbb I ,
 \ee
 with $\mathbb I$ the $n\times n$ identity matrix and
 \be
 \a= \sum_{k=0}^{n-1} \frac1{\sin^2 [\pi(k+x)/n]}=\frac{n^2}{\sin^2 \pi x}. 
 \ee
 Furthermore the traces of the powers of ${\mathbb A}$ have also a particularly simple expression:
 \be
  {\rm Tr}  \mathbb A^{2k}= 
  2(-)^k \sum_{p=1}^{\lfloor n/2\rfloor} (n-2p+1)^{2k}\,.
 \ee
 
It is then natural to expand $\det \mathbb H$ in terms of these traces
 \bea
 \det \mathbb H&=& 
 \det( \a \mathbb I+ \mathbb A^2)= \a^n\det(\mathbb I+\mathbb A^2/\alpha)
 =\a^n\exp[{\rm Tr} \sum_{k=1}^\infty \frac{(-)^{k+1}}{k} \frac{\mathbb A^{2k}}{\a^k} =
 \a^n\exp\left[\sum_{k=1}^\infty \frac{(-)^{k+1}}{\a^k k} {\rm Tr}  \mathbb A^{2k}\right].
\nn
&=&\a^n\exp\left[
-\sum_{p=1}^{\lfloor n/2\rfloor} 2\sum_{k=1}^\infty \frac1k \left(\frac{(n-2p+1)^2}{\a}\right)^k\right]
= \a^n \prod_{p=1}^{\lfloor n/2\rfloor} \exp\left[2 \ln \Big(1-\frac{(n-2p+1)^2}{\a}\Big)\right]
\nn&=& 
\a^n \prod_{p=1}^{\lfloor n/2\rfloor} \left(1- \frac{(n-2p+1)^2}{\a}\right)^2.
\eea 
Finally, using the symmetry for $p\to n+1-p$ and the fact that for $n$ odd the term with $p=(n+1)/2$ is $1$, we have
\bea
F_n^\Upsilon(x)&=& 
\prod_{p=1}^{n} \left(1- \frac{(n-2p+1)^2}{n^2}\sin^2(\pi x) \right)
=\left(\frac{2\sin \pi x}n\right)^{2n}\prod_{p=1}^n\left[\left(\frac{n}{2\sin \pi x}\right)^2-\left(p-\frac{n+1}2\right)^2\right]\\
 &=&\left(\frac{2\sin \pi x}n\right)^{2n}\prod_{p=1}^n 
\left[\left(\frac{n}{2\sin \pi x}\right)-\left(p-\frac{n+1}2\right)\right] \left[\left(\frac{n}{2\sin \pi x}\right)+\left(p-\frac{n+1}2\right)\right].
\nonumber
\eea
Now we can use 
\be
\prod_{p=1}^n(b+p)= \frac{\Gamma(1+b+n)}{\Gamma(1+b)}  ,
\ee
to write
\bea 
F_n^\Upsilon(x)
&=&\left(\frac{2\sin \pi x}n\right)^{2n} (-1)^n
\frac{\Gamma\left(1+\frac12\Big(\frac{n}{\sin\pi x}+n-1\Big)\right)
\Gamma\left(1-\frac12\Big(\frac{n}{\sin\pi x}-n+1\Big)\right)}
{\Gamma\left(1+\frac12\Big(\frac{n}{\sin\pi x}-n-1\Big)\right)
\Gamma\left(1-\frac12\Big(\frac{n}{\sin\pi x}+n+1\Big)\right)}
\nn
&=&  \left(\frac{2\sin (\pi  x)}{n}\right)^{2n}
\frac{ \Gamma^2 \left(\frac{1+n +n \csc (\pi  x) }2 \right)}{
\Gamma^2 \left(\frac{1-n +n \csc (\pi  x) }2 \right)}.
\label{Fnx}
\eea
The result \fr{Fnx} allows us to deduce the following closed form
expressions for the $\rm n^{\rm th}$ R\'enyi entropies
\be
S_n=\frac{c}{6}\Big(1+\frac{1}{n}\Big)\ln\left[\frac{L}{\pi}
\sin\left(\frac{\pi\ell}{L}\right)\right] +\frac{1}{1-n}
\ln\Bigg|
\left(\frac{2\sin (\pi\ell/L)}{n}\right)^{n}
\frac{ \Gamma \left(\frac{1+n +n \csc (\pi\ell/L) }2 \right)}{
\Gamma \left(\frac{1-n +n \csc (\pi\ell/L) }2 \right)}\Bigg|
+c_n +o(L^0).
\label{SnCFT2}
\ee
Furthermore, expanding \fr{Fnx} for small values of $x$ we explicitly recover
Eq. \fr{smallx}. Finally, it is possible to take the derivative at $n=1$  
\be
\left. \frac{\partial F_n^\Upsilon(x)}{\partial n} \right|_{n=1}=
2\left[ \ln\big|2 \sin(\pi x)\big| + \psi \Big(\frac{1}{2\sin(\pi x)}\Big) + \sin(\pi x)\right],
\ee
where we introduced $\psi(z)=\Gamma'(z)/\Gamma(z)$ as the digamma function.
Thus we conclude that the von Neumann entanglement entropy for the Hubbard model in the desired states is 
\be
S_1=
\frac{2}{3}\ln\left[\frac{L}{\pi}
\sin\left(\frac{\pi\ell}{L}\right)\right]-g\Big(\frac{\ell}L\Big) +c_1 +o(1)\ ,
\label{s1cft1}
\ee
where
\be
g(x)=\log\big|2 \sin(\pi x)\big| 
+ \psi\bigg(\frac{1}{2\sin(\pi x)}\bigg) + \sin(\pi x).
\label{gx}
\ee
Third, the above analytic continuation also allows us to extract the limit for $n\to\infty$ of the R\'enyi entropy
which corresponds to the logarithm of the maximum eigenvalue of the reduced density matrix,
also known as single copy entanglement \cite{sce}. 
Taking explicitly the limit from Eq. \fr{Fnx} we obtain the scaling function
\be 
f_\infty(x)=\lim_{n\to \infty} \Big[\frac{1}{2(1-n)} \ln F_n^\Upsilon(x)\Big]
=1+\frac{(1-s) \ln (1-s)-(1+s) \ln (1+s)}{2 s},
\ee
where $s=\sin(\pi x)$.

\begin{figure}[ht]
\includegraphics[width=0.5\textwidth]{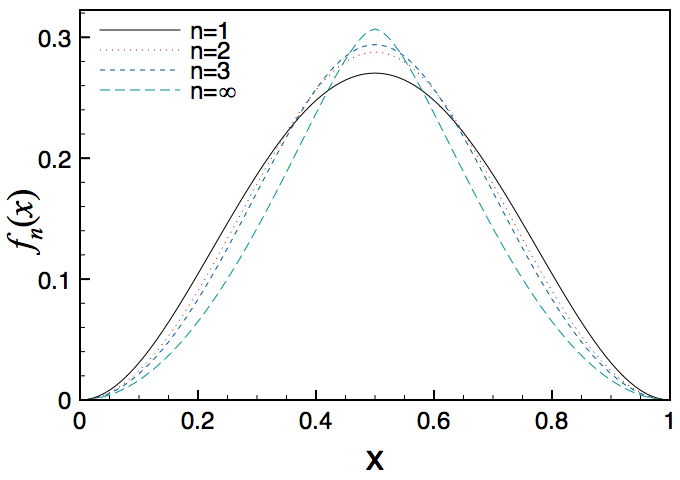}
\caption{The scaling function entering in the R\'enyi entropies $f_n(x)=\frac1{1-n}\ln F_n(x)$ for $n=1,2,3,\infty$ 
as function of $x=\ell/L$.}
\label{Fig:FnCFT}
\end{figure}

In Fig. \ref{Fig:FnCFT} we report the CFT predictions for the scaling
functions of the R\'enyi entropies as function of $x=\ell/L$. 
We note that while for small $x$ the scaling function is smaller for
larger $n$, as given by the expansion \fr{smallx}, for $x=1/2$,
i.e. in the middle of the chain, the behaviour is the opposite.

\subsection{Beyond CFT: effects of the marginal perturbation}
\label{Sec:marg}

As we have already mentioned in Sec. \ref{ssec:Keff}, the low-energy
limit of the Hubbard model gives rise to a Luttinger liquid
Hamiltonian \fr{Hboson}, perturbed by a marginally irrelevant operator
in the spin sector \cite{affleck}. This leads to logarithmic
corrections \cite{logs} in the finite size spectrum and correlation
functions, which can be quite important when trying to compare
analytic results to numerical computations on finite-size systems.
The marginal perturbation is expected to induce logarithmic
corrections to the entanglement entropies, and its effect on
the ground state entanglement in CFTs was studied in some detail in
Ref.~\cite{cc-10}. The corrections to scaling for the ground state
entanglement entropies turn out to be quite small for the
Heisenberg chain\cite{oscillations,chico}, but they turn out to significantly
affect the shell-filling effect in the Hubbard model as we will see
below. A calculation of logarithmic corrections to the entanglement
entropies of the quarter filled Hubbard model is significantly
more difficult than for CFT ground states \cite{cc-10}, and certainly
beyond the scope of our work. Instead, we will attempt to capture
the effects of the marginally irrelevant perturbation
phenomenologically as discussed in Sec. \ref{ssec:Keff}, by replacing
the Luttinger parameter in the spin sector $K_s=1$ by an
\emph{effective} Luttinger parameter $K_s(L,U)$. This leads us to
consider the entanglement in the modified state (\ref{Omod}), rather than
the CFT state \fr{gs4}. The resulting phenomenological scaling
function $F_n^{s}(x)$ can be calculated from the general expression
(\ref{FnO}) with ${\cal O}=\cos\big(\sqrt{2\pi K^{\rm
      eff}_s(U)}\Phi_s(0,0)\big)$. This resulting phenomenological
scaling function $F_2^s(x)$ is of the form
\be
F_2^{s}(x)=\frac{1}{2}
\left[1+\sin^{4K_s}\big(\frac{\pi
    x}{2}\big)+\cos^{4K_s}\big(\frac{\pi x}{2}\big)\right]. 
\label{F2b}
\ee
Analogous formulas for other small integer $n$ can be obtained, but the
lack of a simple determinant representation for general values of $n$
precludes the determination of the corresponding von Neumann entropy.  

\begin{figure}[ht]
\includegraphics[width=0.5\textwidth]{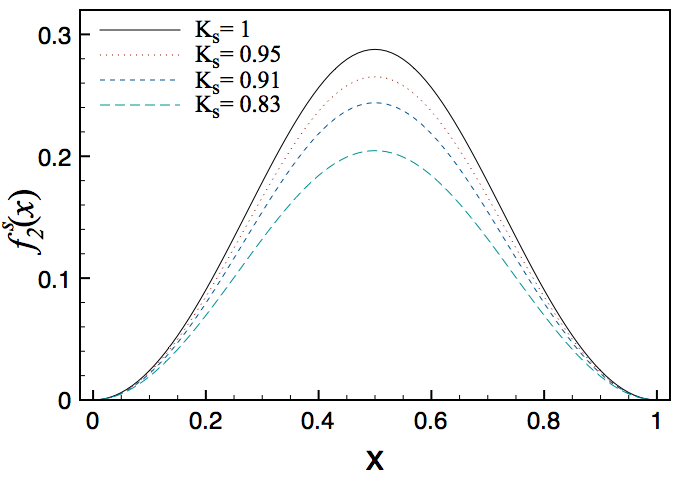}
\caption{The function $f_2^s(x)=-\ln F_2^s(x)$ for
$K_s=1,0.95,0.91,0.83$ as function of $x=\ell/L$. We observe that a
small variation of $K_s$ induces a significant change in
$F_2^s(x)$.} 
\label{Fig:F2k}
\end{figure}
In order to carry out comparisons to numerical results for EEs
we use \fr{F2b} and use the effective Luttinger parameter $K_s^{\rm
  eff}(U)$ determined in Sec. \ref{ssec:Keff} from the finite size
scaling of the ground state energy. 
In Fig. \ref{Fig:F2k} we report the variation of $-\ln F_2^s(x)$ when
$K_s$ varies form $1$ to $1.2$ which is the range of $K_s$ found in
Eq. \fr{table:Ks}. 
\section{Comparison to Numerical Results}
\label{sec:num}

We performed extensive DMRG~\cite{DMRG} computations of the periodic
quarter-filled Hubbard model by keeping $M=3000$ states in order to achieve 
satisfactory convergence for periodic systems up to length $L=64$.
In the following subsections we report the numerical results for several 
coupling parameters and lattice lengths running from $L=8$ to $L=64$,
and covering both sequences of interest, i.e. $L=8n$ and $L=8n+4$. We
perform detailed comparisons of these results with the CFT predictions
obtained in the previous sections.

\subsection{Ground state EEs for $L=4\ {\rm mod}\ 8$.}

\begin{figure}[ht]
\begin{center}
\includegraphics[width=0.45\textwidth]{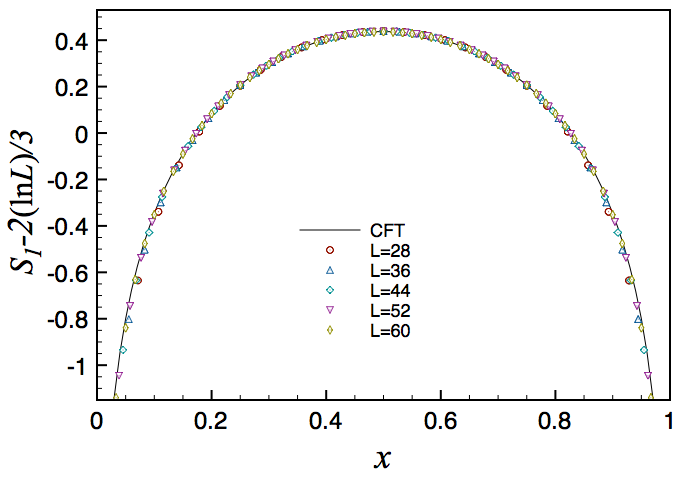}
\includegraphics[width=0.45\textwidth]{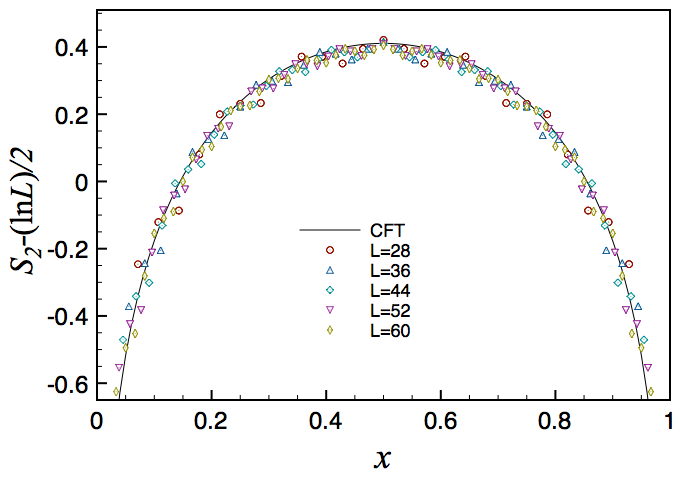}
\end{center}
\caption{Entanglement entropies as a function of $\ell/L$ for $U=4t$ and $L=28,36,44,52,60$.
We report $S_n- 1/3(1+1/n) \ln L$ for $n=1$ (left) and $n=2$ (right). 
The dots are the numerical results and the continuous lines the CFT prediction in Eq. (\ref{CFTan})
plus a non universal additive constant which is fixed by a fit. The agreement is excellent.}
\label{Fig:good1}
\end{figure}

Let us start our analysis from $L=4\ {\rm mod}\ 8$, i.e. the lattice
lengths which should give rise to a standard CFT result
\be
S_n^{\rm CFT}(\ell,L)=
\frac{1}{3}\left[1+\frac{1}{n}\right]\ln\left[\frac{L}{\pi}
\sin\left(\frac{\pi\ell}{L}\right)\right].
\label{CFTan}
\ee
In this case the analysis is quite straightforward: in Fig. \ref{Fig:good1}
we report the entanglement entropies $S_1$ (left panel) and $S_2$ (right panel)
for $U=4t$. 

The data collapse and agreement between numerical data and the CFT
prediction \fr{CFTan} are seen to be very good. In particular, for
$n=1$ no significant deviations are apparent even for moderate lattice
lengths. Conversely for $n=2$, the numerical data oscillate around the
asymptotic result. This is as expected: the oscillations correspond to
the well known ``unusual corrections''  \cite{oscillations,oscillations2,osc2,ot-14,cc-10}
to the scaling of the entanglement entropies, and are known to be
present for all Luttinger liquids and $n\neq1$. We have verified that
these corrections have the structure expected on the basis of the
results obtained in Ref. \cite{oscillations}. Denoting by  
$S_n(\ell,L,U)$ the R\'enyi entropies of the Hubbard model at
interaction strength $U$ on a periodic $L$-site chain, we consider the
scaling of the quantity 
\be
\delta S_n\equiv S_n(\ell,L,U)-S_n^{\rm CFT}(\ell,L),
\label{deltaSn}
\ee
where $S_n^{\rm CFT}(\ell,L)$ is given in Eq. (145). 

\subsection{Ground State EEs for $L=0\ {\rm mod}\ 8$.}
We now turn to lattice lengths $L=8n$, for which we expect a (universal)
$O(1)$ correction of the form \fr{Fnx} to the standard CFT result \fr{CFTan}.
In the following we first consider the von Neumann entropy and the
turn to the second R\'enyi entropy.

\subsubsection{The von Neumann Entropy}
\begin{figure}[ht]
\includegraphics[width=0.325\textwidth]{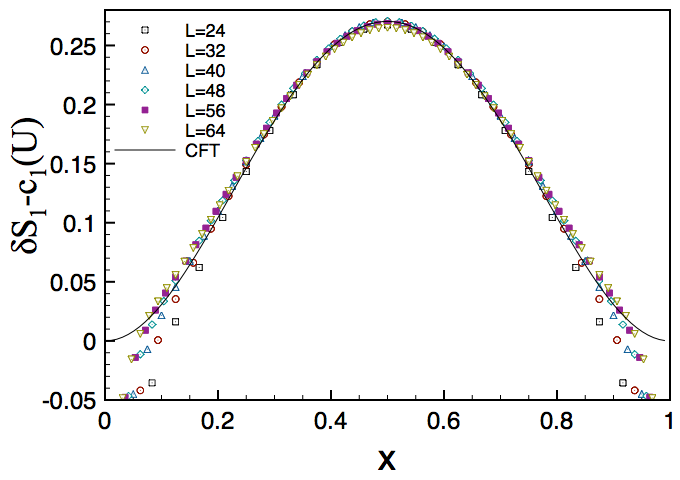}
\includegraphics[width=0.325\textwidth]{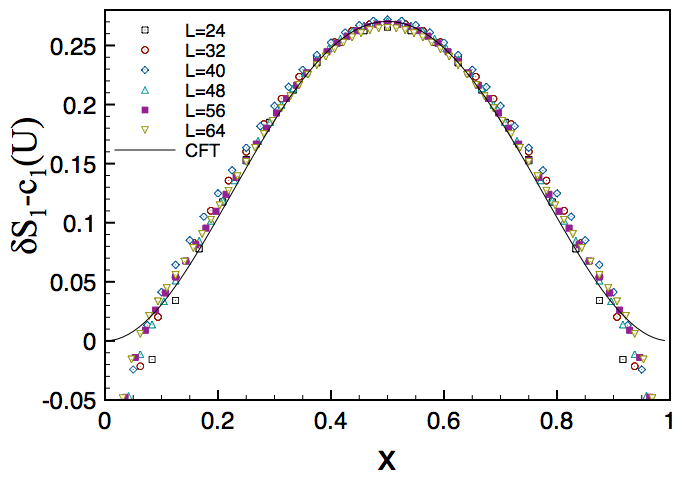}
\includegraphics[width=0.325\textwidth]{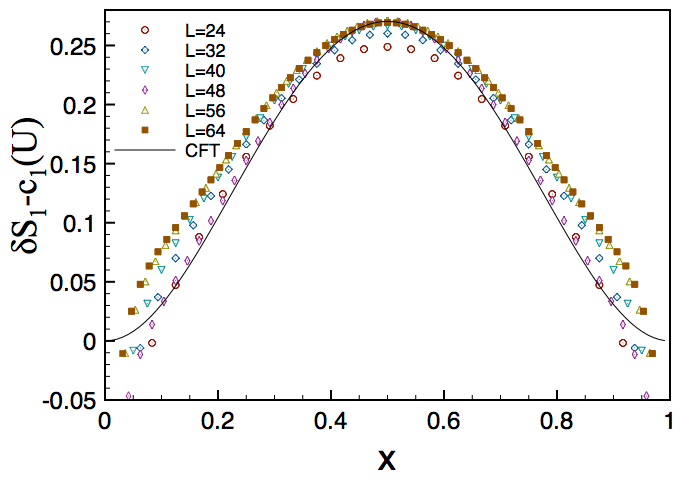}
\caption{
Ground state EEs for $L=8n$. We plot $\delta S_1- c_1(U)$ as a
function of $x=\ell/L$  and $L=24,32,40,48,56,64$. The solid curve is
the CFT prediction $-g(x)$. From left to right the three panels
correspond to $U=0.3t$, $U=t$, and $U=4t$. The non universal constants
are used as fitting parameters. Their values are $c_1(0.3)=1.205$,
$c_1(1)=1.18$, $c_1(4)=1.14$. }
\label{fig:vN1}
\end{figure}

As we have already seen for the case $L=8n +4$,  the von Neumann entropy has
the big advantage compared to the R\'enyi entropies that it does not
contain pronounced oscillatory contributions. Its leading asymptotic behaviour
for large $\ell$ and $L$ is given by the CFT result
\fr{s1cft1}. According to the analysis of Sec.~\ref{sec:CFT}, the
corrections $\delta S_1$ (cf \fr{deltaSn}) for large $L,\ell$
should converge to 
\be
\delta S_1=-g(x) +c_1(U)\ ,
\label{dS1}
\ee
where $g(x)$ is given in Eq. \fr{gx} and $c_1(U)$ is a non-universal
$U$-dependent constant, which we use as a fitting parameter.
The prediction \fr{dS1} is compared to numerical results for $U=0.3t,
t, 4t$ in Fig.~\ref{fig:vN1}. For small values of $U$ $(U=0.3t$) the
agreement is seen to be quite good, but there are increasingly large
deviations when $U$ is increased.

\begin{figure}[ht]
\includegraphics[width=0.32\textwidth]{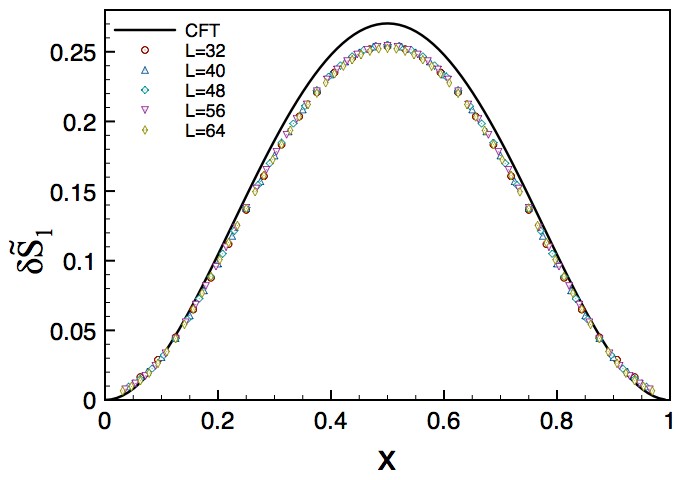}
\includegraphics[width=0.32\textwidth]{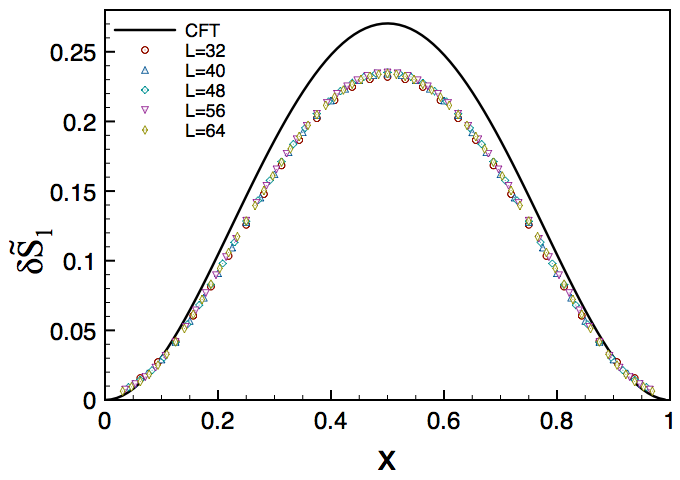}
\includegraphics[width=0.32\textwidth]{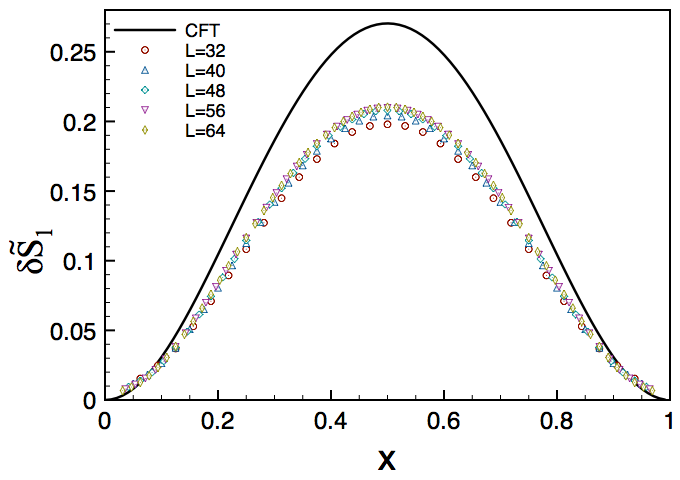}
\caption{$\delta \tilde{S}_1(x,L)$ for $L=32,40,48,56,64$ and
as a function of $x=\ell/L$ for $U=0.3t,t,4t$ from left to right.
The solid line is the asymptotic CFT prediction.  } 
\label{fig:deltavNc}
\end{figure}

An alternative analysis, which eliminates the unknown constants $c_1(U)$,
is based on the fact that the latter are non-universal, but should
not depend on the CFT state considered. In other words $c_1(U)$ is
expected to be the same for EEs calculated in the ground state and in
an excited state of the CFT \cite{abs-11}. In particular, $c_1(U)$
should be the same for $L=8n$ and $L=8n+4$. Hence, by subtracting the
numerical data for $L=8n$ from that for $L=8n+4$, one should directly
obtain the universal CFT function $g(\ell/L)$ (see also
Refs~\cite{abs-11,abs-12}). Implementing such a procedure is not
entirely straightforward, because the accessible values of $\ell/L$
differ for the two sequences of lattice lengths. We circumvent this
problem as follows: since the data for the von Neumann entropy
essentially lie on smooth curves (cf kFigs~\ref{fig:vN1}), we
numerically interpolate the data for a given length in order to obtain
a continuous function $\delta S_1^{\rm int}(x,L)$ with $x\in(0,1)$.
This allows us to compute the difference (here $L=8n$)
\be
\delta \tilde{S}_1(x,L)\equiv\delta S_1(\ell,L)- \delta S_1^{\rm int}(x,L-4).
\label{deltaSt}
\ee
Results for $\delta\tilde{S}_1(x,L)$ are shown in Fig. \ref{fig:deltavNc}. 
We see that the deviations from the CFT prediction $-g(x)$ are still
rather large for $U=t$ and $U=4t$.  

\subsubsection{Origin of the observed deviations: cutoff effects.}

In this section we explore possible sources of the observed deviations
between our numerical results and the CFT prediction \fr{Fnx} in
particular for large values of the interaction strength $U$.

The most basic issue to address is regime of applicability
of the Luttinger liquid description to the Hubbard model. 
Field theory can be used to determine the behaviour of equal time
correlation functions of local operators, as long as the separation
between the latter are sufficiently large. The minimal requirement is
that the separation needs to be large compared to the lattice
spacing. For theories like the Hubbard model the situation is more
complicated, because several degrees of freedom with different
characteristic energy scales are involved. We can estimate the cutoff
in the Luttinger liquid description as follows: we take as an energy
cutoff the bandwidth $W_s$ of the spinon dispersion. At this energy
scale deviations from a linear dispersion in the spin sector are
clearly very large, and lattice effects dominate. We convert this
energy scale to a length scale $\ell_0$ using the charge velocity, as
the latter is always larger than the spin velocity for repulsive
interactions, i.e. 
\be
\ell_0=\frac{v_c}{W_s}.
\ee
The spinon and holon dispersions can be calculated exactly\cite{book},
which leads to the following estimates for $\ell_0$ for a
quarter-filled band
\be
\begin{tabular}{|l|l|l|l|l|}
$U$ & $v_s$ & $v_c$ & $W_s$ & $\ell_0$\\ \hline\hline
0.3 &1.3667 &1.4606 &0.5835 & 2.50\\
\hline
1 &1.2589 &1.5579 &0.5655 & 2.75\\
\hline
4 &0.8818 &1.8148 & 0.4297 & 4.22\\
\hline
16 &0.3444 &1.9789 &0.1721 &11.5\\
\end{tabular}
\ee
The Luttinger liquid description for equal time correlation functions
of local operators is expected to be accurate at length scales large
compared to $\ell_0$, i.e. $x \gg \ell_0$. As far as the finite-size
entanglement entropy is concerned, the relevant length 
scale is the chord distance, and a rough estime for the applicability
of the CFT results to the Hubbard model is then
\be
\frac{L}{\pi}\geq D(\ell,L)\gg \ell_0.
\ee
We see that for increasing $U$ the conformal description is expected
to become worse and eventually ceases to apply for the available
lattice lengths of $L\leq 60$. The upshot of these considerations is
that for very large values of $U$ we should not expect good agreement
between the CFT prediction and numerical results on lattices of
$L\alt 60$ sites.
 
\subsubsection{Origin of the observed deviations: marginally
  irrelevant perturbation.}
A second source for the observed deviations is the presence of the
marginally irrelevant perturbation in the spin sector. An immediate
question that arises in such a scenario, is why this perturbation
should strongly affect the ${\cal O}(1)$ contribtution for lattice
lengths $L=8n$, but appears to be negligible for $L=8n+4$. First, we
in fact expect the marginal perturbation to contribute to the $O(1)$
part of the ground state EEs for $L=8n+4$, but as the latter is
independent of the ratio $\ell/L$ it is quite difficult to spot this
effect numerically. Second, it is known that marginal perturbations
may have much more pronounced effects on corrections to excited state
energies than to the ground state energy\cite{logs}. We conjecture
that the corrections to the entanglement entropies behave in a
similar way.

In order to investigate the possibility that the observed deviations
between the numerical results and the CFT predictions \fr{SnCFT},
\fr{F2-F6} are indeed caused by the marginally irrelevant perturbation
in the spin sector, we now turn to the second R\'enyi entropy and
implement the procedure set out in section \ref{Sec:marg}. In analogy
to the von Neumann entropy, we consider the scaling of the quantity
$\delta S_2$ defined for general $n$ in Eq. \fr{deltaSn}.

\begin{figure}[ht]
\includegraphics[width=0.45\textwidth]{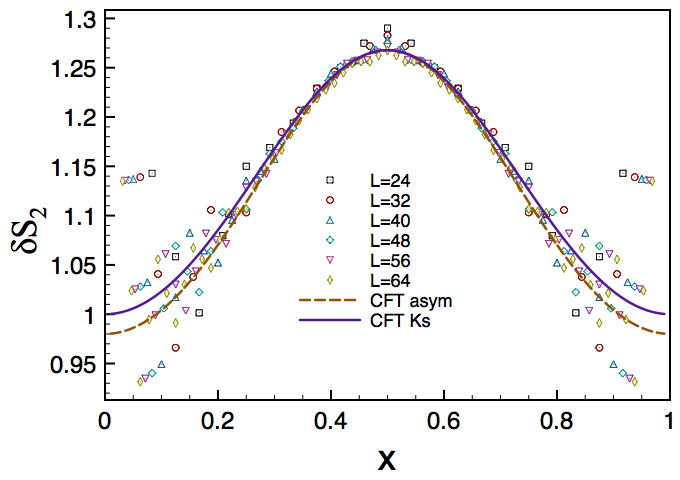}
\includegraphics[width=0.45\textwidth]{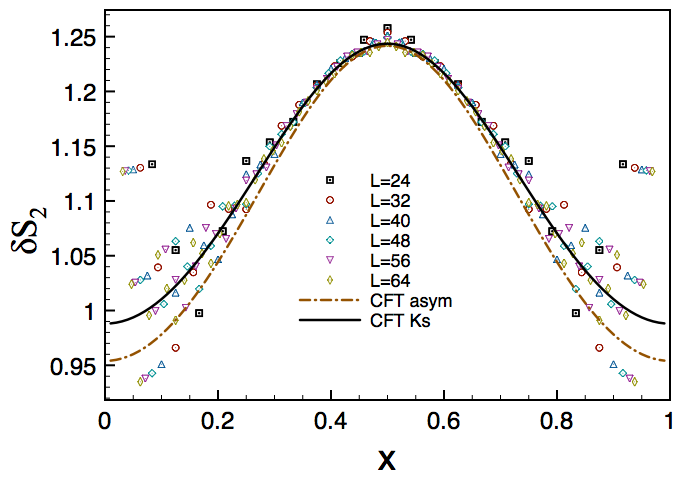}\\
\includegraphics[width=0.45\textwidth]{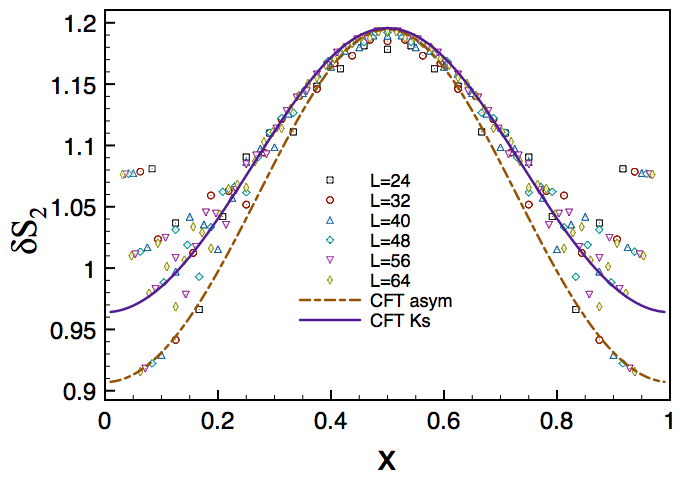}
\includegraphics[width=0.45\textwidth]{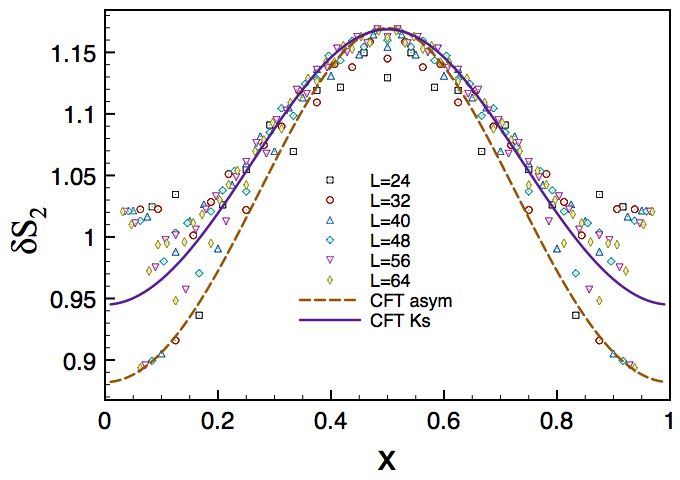}
\caption{Scaling of the second R\'enyi entropy. We plot $\delta S_2$
as a function of $x=\ell/L$ and for $L=24,32,40,48,56,64$. The
different panels correspond to $U=0.3t, t, 4t,16 t$. The dashed line
is the function asymptotic CFT prediction
$c_2(U)-\ln\big(F_2(\ell/L)\big)$. The continuous line is instead the
``effective'' CFT prediction in $c^s_2(U)-\ln\big(F_2^s(\ell/L)\big)$. 
The non-universal constant are used as fit parameters and are taken to
be $c_2(0.3)=0.98$, $c_2(1)=0.954$, $c_2(4)=0.907$, and $c_2(16)=0.882$
in the CFT case, and $c_2^s(0.3)=1.00$, $c_2^s(1)=0.988$,
$c_2^s(4)=0.964$, and $c_2^s(16)=0.945$ in ``effective'' CFT case.
   }
\label{fig:S2a}
\end{figure}

The DMRG data for $\delta S_2(L,\ell)$ for coupling $U=0.3t,t, 4t,
16t$ are reported in Fig.~\ref{fig:S2a} for lattice lengths
$L=24,32,40,48,56,64$. In the various panels of Fig.~\ref{fig:S2a}
the dashed curves correspond to $c_2(U)-\ln\big(F_2(\ell/L)\big)$,
where $c_2(U)$ is fixed by a fit.  
For $U=0.3t$ the agreement is quite good and remaining discrepancies
are compatible with arising from finite-size corrections, i.e. terms
that vanish as $L\to\infty$ at fixed $\ell/L$. However, in analogy to
what observed for the von Neumann entanglement entropy, the agreement
becomes poorer increasing the values of $U$.

As discussed in section \ref{Sec:marg}, the effects of the marginal
irrelevant perturbation on the second R\'enyi entropy can be taken
into account by considering an effective Luttinger parameter $K_s$ in
the spin sector\footnote{As explained in section \ref{Sec:marg} we are
not currently able to implement a similar procedure for the von
Neumann entropy.}. The resulting scaling function is
(cf. Eq. \fr{F2b}) 
\be
F_2^{s}(x)=\frac{1}{2}
\left[1+\sin^{4K_s}\big(\frac{\pi
    x}{2}\big)+\cos^{4K_s}\big(\frac{\pi x}{2}\big)\right]. 
\label{F2b2}
\ee
In absence of renormalization group results we fix the effective
Luttinger parameter $K_s^{\rm eff}(U,L)$ by the {\it independent}
considerations discussed in section \ref{ssec:Keff}, 
Eq. \fr{table:Ks}, in the range of lattice lengths $L$ relevant for
our DMRG computations. More precisely, we compare $\delta
S_2$ defined in Eq. \fr{deltaSn} to 
\be
c_2^s(U)-\ln F_2^{s}\big(\frac{\ell}{L}\big),
\label{F2eff}
\ee
where the constant $c_2^s(U)$ is the only fit parameter since $K_s$ is
fixed in Eq. \fr{table:Ks}. The results of such an analysis are shown
again in Fig. \ref{fig:S2a} as solid lines. We see that the agreement
is now quite satisfactory for {\it all} values of $U$. 

When the interaction strength $U$ is increased from $0$ to $16 t$ for
a $L\sim60$ length lattice, the effective Luttinger parameter
$K_s^{\rm eff}(U,L)$ decreases from $1$ to approximately $0.88$
(cf. Eq. \fr{table:Ks}). This corresponds to a relatively moderate
change of around $15\%$. However, the function $F_2^{s}(x)$ is rather
sensitive to this variation. Recalling the results of Fig.~\ref{Fig:F2k}
we observe that in the middle of the chain  $x=0.5$, the variation in 
$-\ln F_2^{s}(x)$ is approximately $40\%$ when $K_s$ is 
decreased from $K_s=1$ to $K_s=0.88$. This sensitivity of the second
R\'enyi entropy 
to changes in the Luttinger parameter is at the heart of the
significant deviations between the CFT prediction and the numerics.

In our view, the analysis presented above constitutes important
evidence in support of the idea that the main source of disagreement
between the asymptotic CFT calculation and numerical data is the
presence of a marginal irrelevant operator in the spin sector. 

\subsection{Excited State EEs for $L= 4\ {\rm mod}\ 8$}
An key prediction of our theory for the shell-filling effect is that
the additional contribution to the ground state EE for $L=8n$ has
the same functional form as the analogous contribution to a particular
exited state EE for $L=8n+4$. Here we test this prediction by
comparing our CFT results to DMRG compuations. 
The relevant states for $L=8n+4$ are the lowest excitations in the
sector with two added particles and $S^z=0$ and could be either a spin
triplet or a spin singlet, cf Eq. \fr{excgood}. According to the CFT approach
(cf. Sec. \ref{sec:CFT}) their EEs are the same and have the same
functional form as the ground state EEs for $L=8n$. 

According to our CFT prediction, the difference
\be
\Delta S_1^{\rm exc}(\ell,L)=S_1^{\rm exc}(\ell,L)-S_1^{\rm gs}(\ell,L),
\label{DS1}
\ee
for $L=8n+4$ should converge to $-g(x)+c_1(U)$ for $L\to\infty$, where
$g(x)$ is given in \fr{gx}. For small values of $U$ we find good
agreement between this prediction and our numerical results. 

\begin{figure}[ht]
\includegraphics[width=0.4\textwidth]{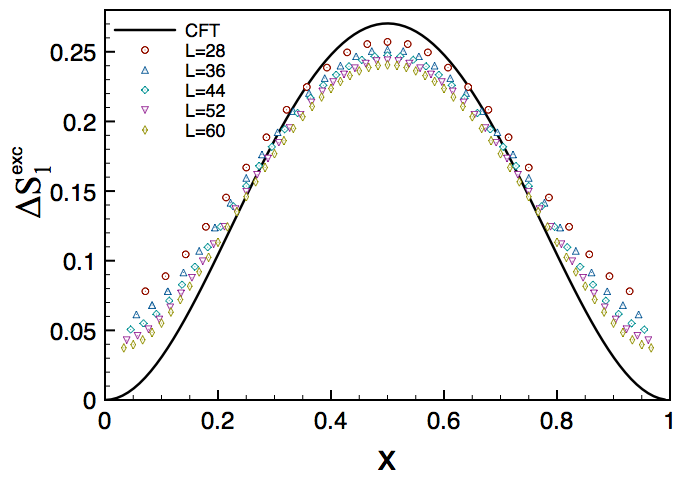}
\includegraphics[width=0.4\textwidth]{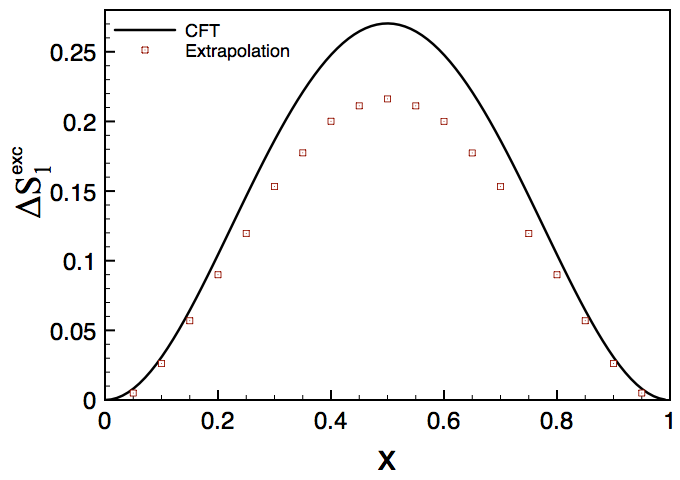}
\caption{Left: $\Delta S_1^{\rm exc}(\ell,L)$ as a function of $\ell/L$ for $U=4t$ and $L=28,36,44,52,60$.   
  Right: Extrapolation of $\Delta S_1^{\rm exc}(\ell,L)$ in the left panel to large $L$. 
  In both panels the solid line is the CFT prediction \fr{s1cft1} for the difference.
  }
\label{fig:difference_good}
\end{figure}

On the other hand, for larger values of $U$ the situation mirrors that
of the ground state EEs for $L=8n$. Results for $\Delta S_1^{\rm exc}$
for $U=4t$ are shown in Fig. \ref{fig:difference_good} (left). The
agreement with the CFT prediction (solid line) at this value of $U$ is
not particularly good, and in addition considerable finite-size
effects are apparent.
This is somewhat surprising, as one might have expected a particularly 
good scaling collapse of the numerical data due to the fact that
\fr{DS1} involves EEs computed for the same lattice lengths. One
reason for the pronounced finite-size effects is that ground and
excited states differ by a shift of the Fermi momentum from
$k_F=\pi/4$ to $\tilde k_F=\pi/4+\pi/L$ as a result of adding two
particles. Plotting \fr{DS1} at fixed $\ell/L$ thus gives rise to
${\cal O}(L^{-1})$ corrections. A possible way to eliminate these,
is to extrapolate the finite-size data. We do this as follows.
We first interpolate the data for a given $L$ to obtain a continuous
function $f_L$ of $x={\ell}/{L}$. We then compute $f_L(x_j)$ for a
selected set of points $x_j\in[0,1]$. Finally we carry out an
extrapolation to infinite system size by fitting a second-order
polynomial in $L^{-1}$ to the sequence $f_L(x_j)$ for a given $x_j$.
The result of this procedure is shown in
Fig.~\ref{fig:difference_good} (right).  We see that there is a clear
discrepancy between the extrapolated numerical and our theoretical
prediction. This plot is quantitatively similar to the one for $U=4t$
in Fig.~\ref{fig:deltavNc} of the ground state EEs for $L=8n$. This
strongly suggests that the origin of the disagreement between the CFT
prediction our numerical results is again the presence of a marginal
irrelevant operator.

\section{Entanglement Entropies in the Extended Hubbard Model}

In the previous section we have accumulated evidence in favour of our
claim, that the observed deviations between EEs in the Hubbard model
and our CFT prediction \fr{SnCFT} are caused by the presence of a
marginally irrelevant perturbation in the spin sector. In order to
remove any lingering doubts, we would like to numerically investigate
a lattice model, for which the coupling constant of the marginally
irrelevant perturbation can be tuned to zero (similar ideas have been
employed in Ref.~\cite{dirk}). This can be achieved by
considering an \emph{extended} quarter filled Hubbard model with
Hamiltonian 
\be
H_{\rm ext}=-t\sum_{j,\sigma}(c^\dagger_{j,\sigma}c_{j+1,\sigma}+ c^\dagger_{j+1,\sigma}c_{j,\sigma})
-\mu\sum_j n_j
+U\sum_j n_{j,\uparrow}\ n_{j,\downarrow}
+\sum_{a=1}^2V_a\sum_j n_jn_{j+a}.
\label{HEHubb}
\ee
This extended Hubbard chain has two additional coupling parameters
$V_1$ and $V_2$ representing density-density interactions between
nearest-neighbour and next-nearest neighbour sites. The model \fr{HEHubb}
is no longer integrable, but as long as $V_1$ and $V_2$ are not too
large compared to $U$, \fr{HEHubb} is known to be in  the same phase
as the Hubbard chain (see \cite{suzu} for the full phase diagram at
small $V_a$ and quarter filling). Crucially, the interactions
$V_{1,2}$ reduce the bare coupling constant $g_{1\perp}$ of the
marginally irrelevant interaction in the spin sector, at least at weak
coupling. In fact, a perturbative calculation gives \cite{suzu}
\be
g_{1\perp}=\frac{Ua_0}{2}-V_2a_0-4D_1a_0^2\big(\frac{U}{2}-V_2\big)
\big(\frac{U}{2}-V_1+V_2\big),
\ee
where $D_1\approx{1.25}/({8\pi ta_0})$. Hence, at weak coupling, the
most efficient way of reducing $g_{1\perp}$ is to take 
\be 
V_2\sim U/2\, .
\ee
We note that the nearest neighbour interaction cannot be used
efficiently to this end, as the linear in $V_1$ contribution vanishes
precisely for quarter filling as the above equation shows. At
low energies and in the parameter regime of interest to us, the model
\fr{HEHubb} is described by a spin-charge separated
Luttinger liquid, perturbed by a marginally irrelevant interaction in
the spin sector, i.e.
\bea
{\cal H}=\sum_{\alpha=c,s}\frac{v_\alpha}{2}\int dx
\left[(\partial_x\Phi_\alpha)^2+ (\partial_x\Theta_\alpha)^2\right]
+\lambda(U,V_2)\int dx
\left[\frac{2}{\pi a_0^2}\cos(\sqrt{8\pi}\Phi_s)+(\partial_x\Theta_s)^2-
(\partial_x\Phi_s)^2\right].
\label{HJJ}
\eea
Increasing $V_2$ from zero (and keeping $V_1=0$ throughout) leads to a
Kosterlitz-Thouless transition at some critical value $V_{2,\rm
  crit}$, which is characterized by  
\be
\lambda(U,V_{2,\rm crit})=0.
\ee
So precisely at $V_{2,\rm crit}$ logarithmic corrections are
absent. Moreover we expect logarithmic corrections to decrease when
$V_2$ is increased from zero to $V_{2,\rm crit}$.

\begin{figure}[ht]
\includegraphics[width=0.32\textwidth]{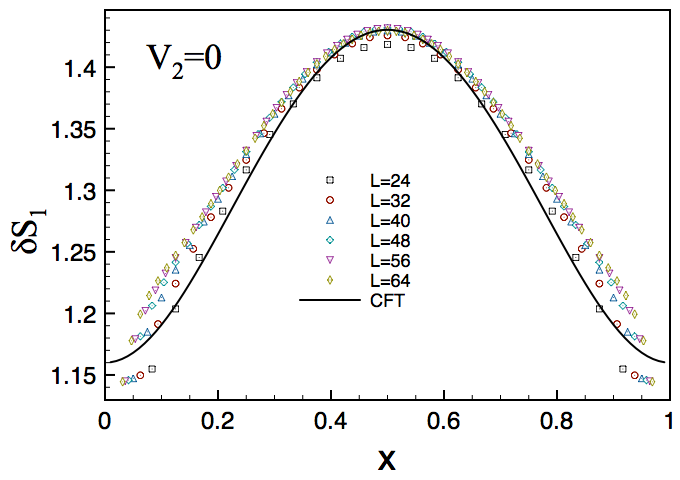}
\includegraphics[width=0.32\textwidth]{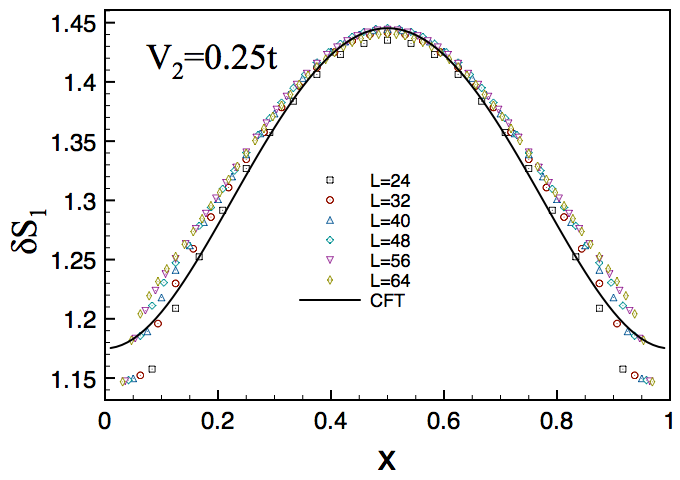}
\includegraphics[width=0.32\textwidth]{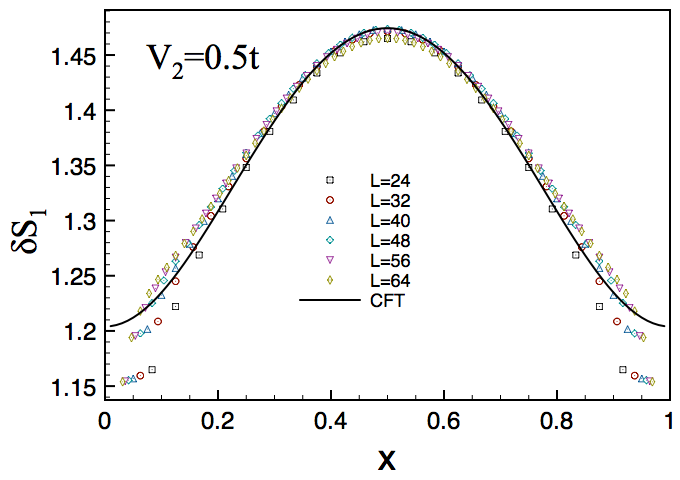}
\caption{$\delta{S}_1$ as a function of $\ell/L$ for $U=2t$ and
  $V_2=0, 0.25 t, 0.5t$ from left to right. The solid line is the CFT prediction \fr{s1cft1}.}
\label{fig:deltaS1V2}
\end{figure}

DMRG results for the subtracted von Neumann entropy $\delta S_1$
(defined in Eq. \fr{deltaSn}) for $U=2t$ and for $V_2=0,0.25t, 0.5 t$
are shown in Fig. \ref{fig:deltaS1V2} and are compared with the
asymptotic CFT formula \fr{s1cft1}.
Clearly the evolution of $\delta S_1$ with $V_2$ is in agreement with
our expectation that the numerical results should approach the
asymptotic CFT prediction as $V_2$ approaches $V_{2,\rm crit}$ from
below.


\section{Conclusions}
\label{concl}
We have presented a detailed analysis of the shell-filling effect in
entanglement entropies of the quarter-filled one-dimensional Hubbard
model with periodic boundary conditions. A short summary of our
results has appeared previously in Ref.~\cite{elc-13}.   
The shell-filling effect, while somewhat unexpected, has a simple
origin: for certain ratios of particle numbers to lattice length, the
ground state in multi-component systems cannot be thought of in 
terms of a simple product of Fermi seas (in general these will consist
of appropriate elementary excitations), but is in fact a linear
combination of products of such seas. 

By means of the Luttinger liquid representation of the Hubbard chain
we developed a CFT approach to calculate the additional ${\cal O}(1)$
contribution to the R\'enyi entropies. These were found to be in very
good agreement with results from DMRG computations for small values of
the interaction strength $U$. For larger values of $U$ we found 
sizeable deviations between the CFT prediction and the DMRG data.
We argued that these deviations can be explained by the presence of a
marginally irrelevant perturbation in the spin sector. In simple
cases such perturbations are known to give rise to logarithmic
corrections to the entanglement entropies \cite{cc-10}.  
We substantiated this view by two complementary considerations. First,
we demonstrated that taking the marginal perturbation into account
semi-phenomenologically through a scale dependent effective Luttinger
parameter in the spin sector leads to a good description of our
numerical data for the second R\'enyi entropy. The analogous
analysis for the von Neumann entropy is presently beyond reach, as the
required analytic continuation in the R\'enyi index remains an open 
problem.
Second, we verified that the shell-filling effect in an
quarter-filled extended Hubbard chain, in which the coupling constant
of the marginally irrelevant perturbation is reduced \cite{suzu}, is
in better agreement with the CFT prediction.

We note that there are similarities between our results and those
for EEs of linear combinations of degenerate ground states
\cite{grover,ben}. However, in our case the ground state is unique for
$U>0$ (and fixed $S^z=0$) and the shell-filling effect does not
a priori require a degeneracy.

The shell-filling effect is a rather general phenomenon as long as
periodic boundary conditions are imposed. It is expected to be present
also for other commensurate fillings in the Hubbard chain,
multi-component continuum or lattice models of interacting fermions or
Fermi-Bose mixtures, and higher dimensional critical systems. Examples
of the former include multi-component gases with delta-function
interactions \cite{Yang-Gaudin} (which has been recently realized
experimentally \cite{leo}), (extended) repulsive $SU(N)$ Hubbard or tJ
models \cite{Schlottmann}. We believe that shell-filling effects may also
come into play in numerical studies of two-dimensional gapless spin
liquids, which display a spinon Fermi surface~\cite{LeeLee,Motrunich,Yang}.

\section*{Acknowledgments} We are grateful to F. Alcaraz and
M. Fagotti  for helpful discussions.  This work was supported by the
EPSRC under grants EP/I032487/1 and EP/J014885/1 (FHLE),  the ERC
under  Starting Grant 279391 EDEQS (PC).

\appendix

\section{Ground state energy for $L=8n$ and logarithmic corrections}
\label{Appa}
In this Appendix we consider the effects of the marginally irrelevant
interaction in the spin sector on the ground state energy.
To that end, we numerically solve the Bethe Ansatz equations for
lattices of up to $L=800$ sites for several values of $U$ and  
calculate the energy of the state \fr{state0}. As a typical example
for $U=4t$ a fit to the ground state energy of the form 
\be
\frac{E}L=\alpha+\beta L^{-2}\ ,
\label{simpleE}
\ee
where we consider the range $448\leq L\leq L=768$, leads to coefficients
\be
\alpha=-0.758043\ ,\quad \beta=1.16479.
\ee
If we only take into account lengths between $L=608$ and $L=768$ we
obtain a slightly 
higher $\beta=1.16995$. The quality of the fit is quite good in either
case: the fit residuals are of order $10^{-9}$ to $10^{-10}$.
Now, if we use a fit function of the form
\be
\frac{E}L=\alpha+\frac{\beta}{L^{2}}+\frac{\gamma}{L^2\log(L/\delta)}\ ,
\label{log}
\ee
we obtain an even better fit (residuals of order $10^{-12}$) with
\be
\alpha=-0.758044\ ,\quad
\beta= 1.34988\ ,\quad
\gamma=1.16939\ ,\quad \delta= 1.56341.
\ee
In the thermodynamic limit we find by solving the integrals equations that
\be
e_0=-0.75804351\ ,\quad
-\pi\frac{v_c+v_s}{6}+\pi v_s=1.3584.
\ee
\emph{This suggests that it is crucial to take the logarithmic
  corrections into account:} the agreement of $\beta$ with the
thermodynamic value is poor if we neglect the logarithmic corrections,
but it becomes quite good if we assume log corrections of the form
\fr{log}. 
The strength of the logarithmic corrections depends on the value of
$U$. Indeed, for $U=0.5t$ a fit to \fr{simpleE} gives
\be
\alpha=-0.871464\ ,\quad\beta=2.56928,
\ee
while a fit of the same data to \fr{log} yields
\be
\alpha=-0.871464\ ,\quad
\beta= 2.68008\ ,\quad
\gamma=1.14495\ ,\quad \delta= 87.5371.
\ee
The residuals are a factor of $10^4$ smaller in the logarithmic fit.
From the solution of the integral equations describing the
thermodynamic limit we have
\be
e_0=-0.87146392\ ,\quad
-\pi\frac{v_c+v_s}{6}+\pi v_s=2.71599.
\ee
The relative error in $\beta$ for the simple fit \fr{simpleE} is $5\%$
for $U=0.5$, but $14\%$ for U=4.

\end{document}